\documentclass[nonblindrev]{informs3_freeuse}

\OneAndAHalfSpacedXI 

\usepackage{natbib}
 \bibpunct[, ]{(}{)}{,}{a}{}{,}%
 \def\bibfont{\small}%

\usepackage{dcolumn}
\usepackage{booktabs}
\usepackage{tikz}
\usetikzlibrary{positioning,shapes,arrows}
\newcolumntype{M}[1]{D{.}{.}{1.#1}}

\usepackage{bm,bbm}
\usepackage{xspace}
\usepackage{enumitem}
\usepackage{mathrsfs}

\newcommand{\tk}{\theta_k}
\newcommand{\pik}{p_{i,k}}
\newcommand{\yik}{y_{i,k}}

\newcommand{\MPSR}{M_{\mbox{\footnotesize PSR}^R}}

\newcommand{\MELFmax}{M_{ELF^{\tilde{R}}}}
\newcommand{\MAELFmax}{M_{ELF^{\tilde{R}}}}
\newcommand{\MELFR}{M_{\mbox{\footnotesize ELF}^{R}}}
\newcommand{\MELF}[1]{M_{\mbox{\footnotesize ELF}^{#1}}}
\newcommand{\MTFSR}{M_{\mbox{\footnotesize ELF}^{R}}}
\newcommand{\MIELFR}{M_{\mbox{\footnotesize I-ELF}^{R}}}
\newcommand{\MASFR}{M_{\mbox{\footnotesize ASF}^{R^{z,z'}}}}

\newcommand{\elf}{ELF\xspace}
\newcommand{\ielf}{I-ELF\xspace}
\newcommand{\tfs}{ELF\xspace}

\newcommand{\E}{\mathop{\mathbf{E}}}

\newcommand{\egcitep}[1]{\citep[e.g.,][]{#1}}
\newcommand{\shortcite}[1]{(\citeyear{#1})}

\TheoremsNumberedThrough     
\ECRepeatTheorems

\EquationsNumberedThrough   

\begin{document}

\RUNAUTHOR{Witkowski et al.}

\RUNTITLE{Incentive-Compatible Forecasting Competitions}

\renewcommand{\thefootnote}{\fnsymbol{footnote}}
\setcounter{footnote}{1}
\TITLE{Incentive-Compatible Forecasting Competitions\footnote{This
    paper is a significantly extended version of
    \cite{witkowski-et-al:2018}.}}

\ARTICLEAUTHORS{%
\AUTHOR{Jens Witkowski}
\AFF{Frankfurt School of Finance \& Management, Frankfurt, Germany, \EMAIL{j.witkowski@fs.de}} 
\AUTHOR{Rupert Freeman}
\AFF{University of Virginia, Charlottesville, VA, USA, \EMAIL{freemanr@darden.virginia.edu}}
\AUTHOR{Jennifer Wortman Vaughan}
\AFF{Microsoft Research, New York, NY, USA, \EMAIL{jenn@microsoft.com}}
\AUTHOR{David M. Pennock}
\AFF{Rutgers University, New Brunswick, NJ, USA, \EMAIL{dpennock@dimacs.rutgers.edu}}
\AUTHOR{Andreas Krause}
\AFF{ETH Zurich, Zurich, Switzerland, \EMAIL{krausea@ethz.ch}}
} 

\ABSTRACT{
  We initiate the study of incentive-compatible forecasting
  competitions in which multiple forecasters make predictions about
  one or more events and compete for a single prize. We have two
  objectives: (1) to incentivize forecasters to report truthfully and
  (2) to award the prize to the most accurate forecaster. Proper
  scoring rules incentivize truthful reporting if all forecasters are
  paid according to their scores. However, incentives become distorted
  if only the best-scoring forecaster wins a prize, since forecasters
  can often increase their probability of having the highest score by
  reporting more extreme beliefs. In this paper, we introduce two
  novel forecasting competition mechanisms. Our first mechanism is
  incentive compatible and guaranteed to select the most accurate
  forecaster with probability higher than any other
  forecaster. Moreover, we show that in the standard single-event,
  two-forecaster setting and under mild technical conditions, no other
  incentive-compatible mechanism selects the most accurate forecaster
  with higher probability. Our second mechanism is incentive
  compatible when forecasters' beliefs are such that information about
  one event does not lead to belief updates on other events, and it
  selects the best forecaster with probability approaching 1 as the
  number of events grows. Our notion of incentive compatibility is
  more general than previous definitions of dominant strategy
  incentive compatibility in that it allows for reports to be
  correlated with the event outcomes. Moreover, our mechanisms are
  easy to implement and can be generalized to the related problems of
  outputting a ranking over forecasters and hiring a forecaster with
  high accuracy on future events.  }%


\maketitle

\section{Introduction}
\label{sec:intro}

\renewcommand{\thefootnote}{\arabic{footnote}}
\setcounter{footnote}{0}

The study of probabilistic predictions dates back to at least the
1950s when meteorologists developed proper scoring rules as a way to
both incentivize truthful forecasts about future events and compare
the relative accuracy of different
forecasters~\citep{brier:1950,good:1952}. Proper scoring rules are
still widely used today to motivate and measure forecasting
accuracy~\egcitep{atanasov-et-al:2017} as well as an active area of
research in decision analysis \egcitep{jose:2017,
  grushka-cockayne-et-al:2017}.

When forecasters are paid their proper scores, they maximize expected
payment by truthfully reporting their beliefs.
However, it is rare to see proper scoring rule payments outside of
experimental labs. Instead, the majority of real-world forecasting
settings are competitions, where forecasters are ranked according to
their score and where prizes are given only to the highest-ranked
forecasters.
Hence, forecasters do not care about maximizing their expected score,
but about whether their forecasts are judged to be better than
others'. For example, in the Good Judgment Project, a recent
geopolitical forecasting tournament, the top 2\% of forecasters were
awarded so-called ``superforecaster''
status~\citep{tetlock-gardner:2015}, which (on top of bragging rights)
gave them full travel reimbursement to a superforecaster
conference.
In play-money prediction markets, forecasters often compete for a
place at the top of a leaderboard~\egcitep{servan-et-al:2004}.
And the same phenomenon holds for algorithmic forecasters; Netflix
offered \$1,000,000 to the team whose machine learning algorithm could
best predict how users would rate movies based on their past
preferences,\footnote{\url{www.netflixprize.com}} and the machine
learning competitions run by Kaggle\footnote{\url{www.kaggle.com}}
rank submitted models based on how well they predict the labels of
data points from an undisclosed test set. One of Kaggle's main uses
today is for recruiters to hire the developers of the best-performing
algorithms~\citep{harris:2013, chakraborty:2016}.

There are good reasons for organizations to run forecasting
competitions as opposed to directly paying each forecaster her proper
score. First, from a marketing perspective awarding a single, large
prize to the winner is more enticing than offering small payments to
everyone. For example, it is unlikely that the Netflix Prize would
have created the same media buzz without offering participants the
prospect of winning \$1,000,000. Second, organizations significantly reduce
transaction costs when only a single or small number of prizes are
awarded. In addition to the literal transaction costs involved in
transferring payments from the organization to the forecasters, there
are sometimes legal reasons that are facilitated by having only a
single transaction.

However, unless they are designed with care, these winner-take-all
competitions can distort incentives, encouraging forecasters to take
big risks as opposed to truthfully reporting their beliefs.
Lichtendahl and Winkler~\shortcite{lichtendahl-winkler:2007} study
a strategic game between two forecasters reporting on a
single event. In their model, each forecaster wishes to maximize her
utility, which is assumed to be a mixture of a proper scoring rule
payment and an (explicit or implicit) bonus for being the best
forecaster,
with a parameter trading off these two components. They show that when
forecasters optimize for their relative rank, they typically want to
report more extreme probabilities than those corresponding to their
true beliefs.

This kind of misreporting is not a purely academic possibility but is
also observed in real-world forecasting competitions. An example is
Kaggle's annual machine learning competition to predict the game
outcomes of the NCAA March Madness college basketball tournament,
where every participant submits up to two statistical models
predicting the outcomes of each possible team pairing. At the end of
the 2017 competition, Andrew Landgraf, the creator of that year's
winning model was interviewed by the Kaggle team about his approach,
saying~\citep{kaggle:2017}: ``\emph{My idea was to model not only the
  probability of each team winning each game, but also the
  competitors’ submissions. Combining these models, I searched for the
  submission with the highest chance of finishing with a prize (top 5
  on the leaderboard). [...] The three main processes are [...]: (1) A
  model of the probability of winning each game, (2) a model of what
  the competitors are likely to submit, and (3) an optimization of my
  submission based on these two models.}''  While rational from a
forecaster's point of view, this strategic behavior creates two
problems for organizations that run forecasting competitions in order
to obtain accurate forecasts: first, the reported forecasts are not
truthful and hence not optimized for accuracy but for ``winning the
game.'' Second, each forecaster responding to the gaming incentives
spends significant effort on strategizing and predicting other
forecasters' behavior instead of investing full effort into acquiring
the most accurate prediction for the event in question.

In this paper, we initiate the study of incentive-compatible
forecasting competitions. After showing that the failure to provide
strict truthfulness incentives is inherent to any deterministic
forecasting competition mechanism, we present the {\bf E}vent {\bf
  L}otteries {\bf F}orecasting Competition Mechanism (\elf). \elf
borrows a trick from the competitive scoring rule of Kilgour and
Gerchak \shortcite{kilgour-gerchak:2004}, which truthfully elicits
probabilistic forecasts for single events. Under Kilgour and Gerchak's
mechanism, a forecaster's payment depends on her relative performance
(measured by a proper scoring rule) compared with other
forecasters. Specifically, her total payment is the difference between
her own score and the average score of all other forecasters. For a
single event, \elf uses a similar idea to compute scores for all
forecasters that are non-negative and sum up to 1. Treating these
scores as a probability distribution over forecasters, \elf then runs
a lottery to determine the winner of the prize. For the prominent
single-event, two-forecaster setting, as also studied by
\cite{lichtendahl-winkler:2007}, we prove that, under mild technical
conditions, there exists no other incentive-compatible mechanism that
selects the more accurate forecaster with higher probability.

Our second mechanism is the Independent-Event Lotteries Forecasting
Competition Mechanism (\ielf), which is specifically designed for
multiple, independent events, and strictly incentive compatible when
forecasters' beliefs are such that information about one event does
not lead to a belief update on the other events. \ielf runs one \elf
lottery for each individual event, eventually awarding the prize to
the forecaster who has won the most event lotteries. As the number of
events grows, \ielf selects the most accurate forecaster with
probability approaching 1. Moreover, both \elf and \ielf are robust
towards unknown risk preferences and our techniques generalize to
other natural settings, such as the incentive-compatible ranking of
forecasters and hiring a forecaster with high accuracy on future
events.

We note here that forecasting
competitions are different from the usual contest settings studied in
the literature, such as innovation contests modeled as all-pay
auctions~\egcitep{konrad:2009}. In those models, while there is also a
prize to be awarded, a participant's strategic choice is the effort
they invest, determining the quality of their provided solution. In
contrast, participants in forecasting competitions strategize about
what they should report given their private information.
Moreover, the mechanism designer's objective is different in the two
settings. Whereas classical contest models seek to maximize the
quality of the provided solutions, the primary objective of
forecasting competitions such as the Good Judgment Project is to
truthfully elicit accurate information from participants.

The question of how to aggregate forecasts has been studied
extensively in the decision analysis community
\egcitep{satopaa-et-al:2014, palley-soll:2019}. We emphasize that
using \elf or \ielf as incentive schemes does not restrict the choice
of whether and how to aggregate forecasts once they have been
elicited. Indeed, a forecasting competition mechanism is not a
substitute for a forecast aggregation algorithm, but a
complement. \cite{lichtendahl-et-al:2013} show that under a
commonly known public-private signal model, a simple average of
``gamed'' forecasts is more accurate than a simple average of truthful
forecasts. However, state-of-the-art aggregation algorithms, such as
the extremized mean \citep{atanasov-et-al:2017} and the logit
aggregator \citep{satopaa-et-al:2014}, consistently outperform simple
averaging in practice and can take advantage of truthful reports.

\section{Model}
\label{sec:model}

We consider a group of $n \geq 2$ forecasters, indexed by
$i \in [n] = \{1, \dots, n\}$, and $m$ events, indexed by
$k \in [m] = \{1, \dots, m\}$. We model these as $m$ random variables
$X_k$ that take values in $\{0,1\}$, and we say that ``event $k$
occurred'' if $X_k = 1$ and that ``event $k$ did not occur'' if
$X_k = 0$.  Independent of the event's outcome, we say that ``event
$k$ materialized.'' Let $\bm{X}$ denote the vector-valued random
variable of event outcomes and
$\bm{x} = \bigl(x_1, \dots, x_k, \dots, x_m\bigr)$ its realization.
Every forecaster $i$ has a subjective belief $\pik \in [0,1]$ of the
probability that event $k$ will occur. We denote the vector of
forecaster $i$'s subjective beliefs over all $m$ events as
$\bm{p}_i = \bigl(p_{i,1}, \dots, \pik, \dots, p_{i,m}\bigr) \in
[0,1]^m.$ All forecasters report their probabilistic forecasts for all
events at the same time, before the first event materializes. (In
Section~\ref{sec:practical-implementation}, we discuss how this
assumption can be relaxed for practical purposes.)
The reported forecast of forecaster $i$ for event $k$ is denoted by
$\yik \in [0,1]$. A forecaster's report can be equal to her true
belief (i.e., $\yik = \pik$) but does not have to be, and we denote
the vector of $i$'s reported forecasts as
$\bm{y}_i = \bigl(y_{i,1}, \dots, \yik, \dots, y_{i,m}\bigr) \in
[0,1]^m.$ In settings with only a single event, i.e., $m=1$, we drop
the subscript $k$ denoting the event from the event outcomes as well
as from the forecasters' reports and beliefs.
Once all $m$ events have materialized, the mechanism selects one of
the $n$ forecasters as the ``winner.'' The selection is based on the
event outcomes and all forecasters' reports on all events. We allow
this selection to be randomized.

\begin{definition}\label{def:selection-scoring-rule}
  A \emph{forecasting competition mechanism} $M$ takes all
  forecasters' reports on all events
  $\bm{y}_1, \dots, \bm{y}_n \in [0,1]^m \times \dots \times [0,1]^m$
  and the materialized outcomes of all events $\bm{x} \in \{0,1\}^m$,
  and selects a forecaster
  $M(\bm{y}_1, \dots, \bm{y}_n,\bm{x}) \in [n]$.
\end{definition}

In contrast to standard proper scoring rules, forecasters only care
about being selected. Every forecaster thus seeks to maximize the
probability of being selected. 
The primary objective is to
incentivize forecasters to report their true beliefs about the
expectation of $\bm{X}$.
Incorporating forecaster $i$'s subjective beliefs, the uncertainty
about other forecasters' reports, and the mechanism's randomization
(if any), we obtain the following definition for strict incentive
compatibility of a mechanism.\footnote{In Section~\ref{sec:accuracy},
  we will introduce a restricted definition that assumes that the
  events $\bm{X}$ are known to be independent and that this
  independence of events is reflected in the uncertainty about others'
  reports.}

\begin{definition}\label{def:truthful}
  Forecasting competition mechanism $M$ is \emph{(robust) strictly
    incentive compatible} if and only if for all forecasters
  $i \in [n]$, all belief vectors $\bm{p}_i$, all joint distributions
  $D$ over outcomes $\bm{X}$ and reports $\bm{Y}_{-i}$ 
  such that the
  marginal distribution of $\bm{X}$ is
  $\mathbf{E}_{\bm{X} \sim D}\bigl[\bm{X}\bigr] = \bm{p}_i$, and all
  alternative report vectors $\bm{y'}_i \neq \bm{p}_i$,
  \begin{equation*}
  \Pr\limits_{\bm{X}, \bm{Y}_{-i} \sim D}\bigl(M(\bm{Y}_1, \dots,
  \bm{p}_i, \dots,
  \bm{Y}_n,\bm{X}) = i\bigr) > \Pr\limits_{\bm{X}, \bm{Y}_{-i} \sim D}\bigl(M(\bm{Y}_1, \dots,
  \bm{y'}_i, \dots,
  \bm{Y}_n,\bm{X}) = i\bigr).
\end{equation*}
\end{definition}
Observe that this definition of incentive compatibility is very
general, allowing us to capture, for instance, that forecaster $i$
believes that $j \neq i$ perfectly forecasts the correct outcome while
$i$ herself does not. More generally, it allows for settings in which
forecaster $i$ would update her belief upon learning forecaster $j$'s
report. In particular, our definition of incentive compatibility
applies to standard Bayesian models, where the participants' beliefs
stem from noisy observations of some ground
truth~\egcitep{lichtendahl-winkler:2007}. 
This is in contrast to
previous work that defined \emph{immutable-belief incentive
  compatibility}~\citep{kilgour-gerchak:2004,lambert-et-al:2008a},
which only requires truthful reporting to be optimal when the reports
of other forecasters are constant (i.e., with no dependence on each
other or the event outcomes). We refer the reader to
Appendix~\ref{sec:bayesian-model} for an extensive discussion of this
distinction, which also includes a concrete numerical example showing
that immutable-belief incentive compatible mechanisms suggested in the
literature tend to incentivize misreports in Bayesian
contexts.
In contrast to previously studied competitive forecasting settings,
most notably those by \cite{lichtendahl-winkler:2007} and
\cite{lichtendahl-et-al:2013}, we do not require $\bm{p}_i$ to come
from any particular parametric belief model. Moreover, and crucially,
we do not restrict our analysis to Bayes' Nash equilibrium
play. Instead, and in line with the literature on (single-forecaster)
proper scoring rules~\egcitep{gneiting-raftery:2007}, the mechanisms
we design obtain strict incentive compatibility in dominant
strategies. That is, our objective is to provide strict incentives for
truthful reports independent of the reports of other
forecasters.

Also observe that we do not require the typical assumption that
forecasters are risk neutral: every forecaster strictly prefers being
selected over not being selected, so that the higher the probability
of being selected, the better. This idea is not new; previous work
used lotteries to address unknown risk preferences of
forecasters~\citep{karni:2009,lambert:2011,hossain-okui:2013}. While
we also reward forecasters probabilistically (and obtain robustness to
unknown risk preferences as a bonus\footnote{In fact,
    we do not even require that forecasters are expected utility
    maximizers but only require that they are ``probabilistically
    sophisticated''~\citep{machina-schmeidler:1992}. We thank an
    anonymous reviewer for this observation.}), the primary reason we use
lotteries is because we have many forecasters but only a single prize
to award. To the best of our knowledge, we are the first to study this
competitive lottery setting in the context of forecasting.

\section{Forecasting Competitions Using Standard Proper Scoring Rules}
\label{sec:psr}

Consider a single forecaster and a single event $X$. A scoring rule
$R$ computes a payment that depends on the materialized event outcome
$x$ and the forecaster's report $y \in [0,1]$ regarding the
probability that $X = 1$, paying the forecaster some amount $R(y,x)$.

\begin{definition}[Strictly Proper Scoring Rule]\label{def:proper}
  A \emph{scoring rule} $R$ is a mapping from reports $y \in [0,1]$
  and outcomes $x \in \{0,1\}$ to scores
  $R(y,x) \in \mathbb{R}\, \cup\, \{-\infty\}$. A scoring rule $R$ is
  \emph{strictly proper} if, for all $y, p \in [0,1]$ with $y \neq p$,
  it holds that
  $\E_{X \sim p} \bigl[R(p,X)\bigr] > \E_{X \sim
    p}\bigl[R(y,X)\bigr]$. $R$ is \emph{bounded} if there exist
  $\underline{R}, \overline{R} \in \mathbb{R}$ such that
  $R(y,x) \in [\underline{R}, \overline{R}]$ for all
  $y \in [0,1] ,\, x \in \{0,1\}$. Proper scoring rule $R$ is
  \emph{normalized} if it is bounded between $0$ and $1$, and if
  $R(0,0) = R(1,1) = 1$ and $R(y,x) = 0$ for some $y \in [0,1]$ and
  $x \in \{0,1\}$.
\end{definition}

When clear from context, we will write $R \in [0,1]$ to refer to a
scoring rule bounded between 0 and 1.
There exist infinitely many proper scoring rules since any (strictly)
convex function yields a (strictly) proper scoring
rule~(\citeauthor{gneiting-raftery:2007},
\citeyear{gneiting-raftery:2007}; Theorem 1).
A widely used bounded scoring rule is the \emph{quadratic scoring
  rule}~\citep{brier:1950}, which we will regularly refer to
throughout the paper and give here in its canonical, normalized form.
\begin{proposition}\label{thm:qsr}
  \citep{brier:1950} The \emph{quadratic scoring rule}
  $R_q(y,x) = 1-(y-x)^2$ is strictly proper.
\end{proposition}

Bounded proper scoring rules used in practice are often already
normalized. For example, both the quadratic scoring rule and the
spherical rule~\egcitep{jose:2009} already are.
We note that any bounded proper scoring rule $R$ can be transformed
into a normalized proper scoring rule
$\tilde{R}$,
and refer the reader to Appendix~\ref{sec:normalization} for details.

\subsection{Mechanism}\label{sec:mpsr}

A natural way to extend a strictly proper scoring rule $R$ to a
forecasting competition mechanism is to output the forecaster with
highest score according to $R$, summed across all $m$ events. This
mechanism is commonly used in practice to select top forecasters,
including by the Good Judgment Project~\citep{tetlock-gardner:2015}
and FiveThirtyEight's NFL Forecasting
Game.\footnote{\url{https://projects.fivethirtyeight.com/2019-nfl-forecasting-game}}
Let $\MPSR$ denote the mechanism derived in this way from proper
scoring rule $R$. That is, $\MPSR$ selects the forecasters with
highest score,
\[ \MPSR(\bm{y}_1, \ldots, \bm{y}_n, \bm{x}) \in \argmax_{i \in [n]}
  \sum_{k=1}^m R(\yik, x_k) ,\] with ties broken by forecaster
index.\footnote{Other tie-breaking procedures are possible and our
  results do not rely on any particular one.}

\subsection{Incentive Analysis}

It is well known that selecting a forecaster according to highest
proper scoring rule score may produce perverse incentives. In general,
forecasters are incentivized to make over-confident reports to
increase their chance of being judged the best forecaster \emph{ex
  post} for at least some outcomes. To see this, consider an event $X$
and two forecasters who believe that $X$ occurs with probability 0.8
and 0.9, respectively. If both report their beliefs truthfully, the
forecaster who reports 0.8 achieves the highest score---and is
therefore selected by the mechanism---whenever $X=0$, which she
believes to occur with probability 0.2. However, if she instead
reports some $y>0.9$, she is selected by the mechanism whenever $X=1$,
which she believes to occurs with probability 0.8.
We present a more general example illustrating the failure of
incentive compatibility of proper scoring rule selection for any
$n \ge 2$ and $m \ge 1$ in Appendix~\ref{sec:mbr-fails}. For a
thorough analysis of the (non-truthful) strategic behavior of
competitive forecasters when ranked by standard proper scoring rules,
we defer to \cite{lichtendahl-winkler:2007}.
Moreover, as shown by Theorem~\ref{thm:deterministic-not-truthful},
failure to provide strict incentive compatibility is inherent to any
deterministic forecasting competition mechanism. For intuition, the
proof proceeds by showing that any deterministic mechanism only has
finitely many possible outputs, whereas each agent has an infinite
reporting space, and hence, forecasters cannot always strictly prefer
truthful reporting.

\begin{theorem} \label{thm:deterministic-not-truthful} No
  deterministic forecasting competition mechanism is strictly
  incentive compatible.
\end{theorem}

\section{Incentive-Compatible Forecasting Competitions}
\label{sec:single-event}

Theorem~\ref{thm:deterministic-not-truthful} motivates the study of
randomized forecasting competition mechanisms. In
Section~\ref{sec:mechanism-se}, to build intuition, we begin by
considering the single-event setting ($m=1$) and introduce the Event
Lotteries Forecasting Competition Mechanism (\elf), a strictly
incentive-compatible mechanism. In
Section~\ref{sec:correlation-proof-mech}, we then show how to extend
\elf to handle multiple, arbitrarily correlated events.

What needs to hold in order for a forecasting competition to be
strictly incentive compatible? First note that strict incentive
compatibility requires that, for any beliefs over outcomes $X$ and
reports $Y_{-i}$, the probability $f_i$ of selecting forecaster $i$
must behave like a strictly proper scoring rule for $i$. If this is
not the case, then $i$ could increase her probability of being
selected by misreporting.
Thus, we need strictly proper scoring rules for each forecaster that
are non-negative and always sum to 1 so that they form a valid
probability distribution. A natural first attempt to achieve this
would be to use any strictly proper scoring rule, such as the
quadratic scoring rule $R_q$, and ``normalize'' by dividing by the sum
of all forecasters' scores.
However, such a \emph{multiplicative} normalization violates incentive
compatibility because the factor by which scores are normalized is
1/(sum of forecasters' scores), which may differ between outcomes,
causing forecasters to bias their predictions towards less likely
outcomes. For an example illustrating this phenomenon, see
Appendix~\ref{sec:strawman-fails}.

To get around this, we borrow a trick from the competitive scoring
rule mechanism of Kilgour and
Gerchak~\shortcite{kilgour-gerchak:2004}, which takes advantage of the
fact that incentive compatibility is preserved when adding or
subtracting a function of other reports and the outcome.
Using their mechanism, each forecaster's payment is her score
according to a proper scoring rule minus the average score of all
other forecasters.
Our Event Lotteries Forecasting Competition Mechanism (\elf) uses a
similar idea to normalize all forecasters' scores
\emph{additively}, so that they are non-negative and sum up to 1. \elf
then runs a lottery based on these scores to determine the winner of
the prize.

\subsection{Single-Event Mechanism}\label{sec:mechanism-se}

For a single event, the {\em Event Lotteries Forecasting Competition
  Mechanism (\elf)} $\MELFR(y_1, \dots, y_n,x)$ selects forecaster
$i \in [n]$ with probability
\begin{equation}\label{eq:se-elf}
  f_i(y_1, \dots, y_n,x) = \frac{1}{n} + \frac{1}{n} \biggl(R\bigl(y_i,x\bigr)-\frac{1}{n-1} \sum_{j
    \not= i} R\bigl(y_j,x\bigr)\biggr),
\end{equation}
where $R \in [0,1]$ is a bounded strictly proper scoring
rule.\footnote{Although our definition allows for any bounded $R$, we
  will see in Section~\ref{sec:accuracy} that the optimal accuracy
  guarantees are achieved for normalized $R$.}

One can think of \elf as giving each forecaster a $1/n$ probability to
start with, adjusting this up or down depending on how their
performance compares to that of other forecasters.
It is easy to see that the vector\footnote{We drop the dependencies of
  each $f_i$ for clarity.} $\bigl(f_1, \dots, f_n\bigr)$ is a valid
probability distribution: that each $f_i$ is non-negative follows
immediately from $R$ being bounded in $[0,1]$,
and $\sum_{i=1}^n f_i = 1$ since
\[
  \sum_{i=1}^n f_i = 1+ \frac{1}{n}\sum_{i=1}^n
  \biggl(R\bigl(y_i,x\bigr)-\frac{1}{n-1} \sum_{j \neq i}
  R\bigl(y_j,x\bigr)\biggr) = 1+ \frac{1}{n} \biggl( \sum_{i=1}^n
  R\bigl(y_i,x\bigr)-\frac{n-1}{n-1} \sum_{i=1}^n
  R\bigl(y_i,x\bigr)\biggr)=1.
\]

Generalizing the result of Kilgour and
Gerchak~\shortcite{kilgour-gerchak:2004} to incorporate Bayesian
reasoning about other forecasters, we can show that \elf is incentive
compatible.

\begin{theorem}\label{thm:single-event-truthfulness}
  The Event Lotteries Forecasting Competition Mechanism $\MELFR$ is
  strictly incentive compatible for $m=1$.
\end{theorem}

\subsection{Multiple-Event Mechanism}\label{sec:correlation-proof-mech}

We now consider a natural generalization of single-event \elf to
multiple events. For multiple events, \tfs proceeds as follows after all events have
materialized.
$\MTFSR(\bm{y}_1, \ldots, \bm{y}_n,\bm{x})$ selects forecaster
$i \in [n]$ with probability
\begin{equation}\label{eq:tfs}
g_i(\bm{y}_1, \ldots, \bm{y}_n,\bm{x}) = \frac{1}{m}\sum_{k=1}^m f_{i,k} ,\,\, \text{where}\,
f_{i,k}= \frac{1}{n} + \frac{1}{n}
\biggl(R\bigl(\yik,x_k\bigr)-\frac{1}{n-1} \sum_{j \not= i}
R\bigl(y_{j,k},x_k\bigr)\biggr),
\end{equation}
and where $R \in [0,1]$ is a bounded strictly proper scoring rule.

This corresponds to running single-event \elf for every event, and
selecting each forecaster with probability equal to the average
probability assigned to her across all events. Note that this
procedure can equivalently be interpreted as sampling a single event
uniformly at random, and awarding the prize to the forecaster selected
by single-event \elf on that event.
Strict incentive compatibility of \tfs then follows directly from strict
incentive compatibility of single-event \elf.

\begin{theorem}\label{thm:a-elf-truthfulness}
  The Event Lotteries Forecasting Competition Mechanism $\MELFR$ is
  strictly incentive compatible for $m \ge 1$ events.
\end{theorem}

\section{Incentive-Compatible and Accurate Forecasting Competitions}
\label{sec:accuracy}

The \tfs mechanism from Section~\ref{sec:correlation-proof-mech} is
strictly incentive compatible for arbitrarily correlated events. If
(strict) incentive compatibility is the only objective, \tfs provides
a definitive solution. In many settings, however, the system designer
strives for an additional objective, namely that the prize is awarded
to the most accurate forecaster. 
In the Good Judgment Project, for
example, the 2\% of forecasters with highest quadratic scores were
awarded ``superforecaster'' status~\citep{tetlock-gardner:2015}. It is
implicit in the term that these individuals should be the most
accurate forecasters. Similarly, recruiters on Kaggle seek to make job
offers to the data scientists who create the most accurate
models~\citep{harris:2013}.
Hence, in addition to incentive
compatibility, the objective in this work is to select the forecaster
with the highest accuracy with as high a probability as possible, and
ideally with probability tending to 1 as the number of events grows.
Of course, one could imagine other objectives, such as
maximizing the expected accuracy of the selected forecaster or
minimizing the accuracy gap between the selected and the best
forecaster. We briefly discuss alternatives in
Section~\ref{sec:discussion}.

In judging accuracy, one needs to have a model for ground truth. Here,
we borrow from statistical learning theory and assume that event
outcomes are drawn from an unknown joint probability distribution
$\theta$ over $X_1, \dots, X_m$.
We emphasize that $\theta$ is latent and hence never observed by either the forecasters or the mechanism.
The marginal probability that event $k$ will occur is denoted by
$\tk \in [0,1]$. Note that this is strictly more general than defining
outcomes as ground truth since, in particular, it allows for
$\theta_k = x_k$.
In Definition~\ref{def:proper}, proper scoring rules are defined in an
incentive spirit, as a tool for the incentive-compatible elicitation
of subjective beliefs. In particular, the expectation is taken with
respect to a forecaster's subjective belief $p$. Proper scoring rules
also have an accuracy interpretation.
If the expectation is taken with respect to the true probability
$\theta_k$ of event $k$ occurring, then properness implies that
reporting the true probability obtains a higher expected score than
any other report. Reports that do not coincide with the true
probability lead to lower expected scores, and different proper
scoring rules correspond to different accuracy measures in that they
punish reports diverging from the true probability differently.
For example, with true probability $\theta_k$, the quadratic scoring
rule (Proposition~\ref{thm:qsr}) punishes a report $y$ by
$\E_{X_k \sim \theta_k}\bigl[R_q(\theta_k,X_k) - R_q(y,X_k)\bigr] =
\bigl(y-\theta_k\bigr)^2$.

Importantly, the choice of proper scoring rule has implications for
the relative rank of forecasters. For example, let $\theta_k = 0.7$
and let two forecasters report $y_{1,k} = 0.9$ and $y_{2,k} = 0.51$,
respectively. Then, under the quadratic scoring rule, forecaster 2
obtains a higher expected score than forecaster 1 (less punishment),
whereas under the spherical scoring rule,\footnote{The spherical
  scoring rule~\citep{jose:2009} is defined as $R_s(y,x) :=
  \frac{yx+(1-y)(1-x)} {\sqrt{y^2+(1-y)^2}}$. Forecaster 1 obtains an
  expected score of 0.73 and forecaster 2 obtains an expected score of
  only 0.71.} forecaster 1 obtains a higher expected score than
forecaster 2.
That is, the system designer's choice of proper scoring rule in a
forecasting competition determines the (relative) accuracy measure
that forecasters are judged by. For the incentive-compatible
mechanisms in this paper, the proper scoring rules need to be
bounded. In particular, the accuracy measure implied by the unbounded
logarithmic scoring rule~\citep{good:1952} cannot be used. Note that
this restriction to bounded scoring rules (such as the quadratic or
spherical scoring rule) is also present outside of competition
settings when forecasters are simply paid their score as one cannot
ensure non-negative payments for unbounded scoring
rules. Moreover, we will later show in
  Theorem~\ref{thm:unbounded-impossibility} that no other
  incentive-compatible forecasting competition mechanism can implement
  accuracy measures corresponding to unbounded scoring rules under
  mild technical assumptions.
Hence, for the remainder of the paper (with the exception of
Theorem~\ref{thm:unbounded-impossibility}), the accuracy measure that
is used will be given by a particular bounded proper scoring rule. The
objective will be to select the forecaster with highest expected score
according to that scoring rule while ensuring that the mechanism is
strictly incentive compatible even in the competition setting.
For this, it is helpful to overload notation of proper scoring rule
$R$ and define
\begin{equation*}
R(\bm{y}_i, \theta) := \E\limits_{\bm{X} \sim \theta} \frac{1}{m} \sum_{k=1}^m R(y_{i,k},X_k)
\end{equation*}
as the expected score of report $\bm{y}_i$ using $R$ and given joint
probability $\theta$. This allows us to make statements about the
relative accuracy of forecasters with respect to $R$ and $\theta$. In
particular, forecaster $i$ is more accurate than forecaster $j$ on the
$m \geq 1$ events if and only if $R(\bm{y}_i, \theta) > R(\bm{y}_j,
\theta)$.

\subsection{Accuracy of \elf}

We first observe that \tfs selects forecasters with
higher accuracy more often than those with lower accuracy.
\begin{definition}
  Forecasting competition mechanism $M$ is \emph{rank accurate} with
  respect to proper scoring rule $R$ if and only if it holds that
  $R(\bm{y}_i,\theta) > R(\bm{y}_j,\theta) \Leftrightarrow
  \Pr\limits_{\bm{X} \sim \theta}\Bigl(M\bigl(\bm{y}_1, \dots,
  \bm{y}_n,\bm{X}\bigr) = i\Bigr) > \Pr\limits_{\bm{X} \sim
    \theta}\Bigl(M\bigl(\bm{y}_1, \dots, \bm{y}_n,\bm{X}\bigr) =
  j\Bigr)$ for all joint distributions $\theta$ over
  $X_1, \dots, X_m$, all $\bm{y}_1, \dots, \bm{y}_n \in [0,1]^m$, and
  all $i,j \in [n]$.
\end{definition}
The next statement follows immediately from taking expectation over
$\bm{X}$ in Equation~\ref{eq:tfs}.

\begin{proposition} \label{prop:single-event-accuracy}
  The probability that $\MTFSR$ selects forecaster $i$ given joint
  probability $\theta$ is
  $\Pr\limits_{\bm{X} \sim \theta}\Bigl(\MTFSR\bigl(\bm{y}_1, \dots,
  \bm{y}_n,\bm{X}\bigr) = i\Bigr) =
  \frac{1}{n}+\frac{1}{n}\left(R(\bm{y}_i, \theta) - \frac{1}{n-1}
    \sum_{j \not= i} R(\bm{y}_j, \theta)\right)$.
\end{proposition}

\begin{corollary}
	\label{cor:rank-accuracy}
  $\MTFSR$ is rank accurate with respect to $R$. In particular, it
  selects the most accurate forecaster with higher probability than
  any other forecaster.
\end{corollary}

One may wonder if there exist incentive-compatible forecasting
competition mechanisms that select the most accurate forecaster with
higher probability than \elf. In Theorem~\ref{thm:cant-do-better} we
rule out this possibility for the standard two-forecaster, single-event
setting~\egcitep{lichtendahl-winkler:2007}, subject to mild conditions
on the form of the forecasting competition mechanism.

\begin{definition}
Forecasting competition mechanism $M$ is \emph{anonymous} if the selected
forecaster does not depend on the identities of the forecasters. That
is, $M$ is anonymous if for any permutation $\sigma$ of $[n]$, any
forecaster $i$, any reports $\bm{y}_1, \dots, \bm{y}_n$, and any
outcome vector $\bm{x}$, it holds that
$\Pr\bigl(M(\bm{y}_1, \dots, \bm{y}_n,\bm{x}) = i\bigr) =
\Pr\Bigl(M(\bm{y}_{\sigma^{-1}(1)}, \dots,
\bm{y}_{\sigma^{-1}(n)},\bm{x}) = \sigma(i)\Bigr)$.
\end{definition}

In order to exploit existing characterization theorems of competitive
scoring rules~\citep{lambert-et-al:2008a}, we restrict attention to smooth
forecasting competition mechanisms in Theorem~\ref{thm:cant-do-better}.

\begin{definition}
A forecasting competition mechanism $M$ is \emph{smooth} if the
corresponding function that outputs a probability distribution over
forecasters, $\Pr\bigl(M(\bm{y}_1, \dots, \bm{y}_n,\bm{x})\bigr)$, is
twice continuously differentiable with respect to each $\bm{y_i}$.
\end{definition}

Theorem~\ref{thm:cant-do-better} shows that if a strictly
incentive-compatible mechanism $M$ ever selects the more accurate
forecaster from a single-event, two-forecaster competition with higher
probability than \elf with normalized $R$, then $M$ is not rank
accurate with respect to $R$, i.e., there must exist another instance
in which $M$ selects the less accurate forecaster with higher
probability than the more accurate one.  Recall that we denote by
$\tilde{R}$ the proper scoring rule that results from normalizing $R$
as described in Appendix A.

\begin{theorem}
  \label{thm:cant-do-better}
  Let $M$ be a smooth and anonymous forecasting competition mechanism
  that is rank accurate with respect to $R$ and for which there exist
  $y_1, y_2 \in [0,1]$ and distribution $\theta$ such that $R(y_1,
  \theta)>R(y_2,\theta)$ and $\Pr\limits_{X \sim \theta}\bigl(M(y_1,
  y_2,X) = 1\bigr) > \Pr\limits_{X \sim \theta}\bigl(\MAELFmax(y_1,
  y_2,X) = 1\bigr)$. Then $M$ is not strictly incentive compatible.
\end{theorem}

By adapting elements of the proof of Theorem~\ref{thm:cant-do-better},
we obtain an impossibility result for unbounded scoring rules.
\begin{theorem}
  \label{thm:unbounded-impossibility}
  Let $R$ be an unbounded scoring rule. No smooth, anonymous, and
  strictly incentive compatible forecasting competition mechanism is
  rank accurate with respect to $R$.
\end{theorem}
A notable consequence of Theorem~\ref{thm:unbounded-impossibility}
concerns the logarithmic scoring rule, which is the proper scoring
rule most grounded in classical information
theory~\citep[e.g.,][Section~2.2]{gneiting-raftery:2007}. In
particular, the theorem implies that no incentive compatible
forecasting competition mechanism is rank accurate with respect to the
logarithmic rule.

\subsection{Accuracy in the Limit}\label{sec:limi-accuracy}
  
Theorem~\ref{thm:cant-do-better} shows that we cannot do better than
\elf for the standard single-event, two-forecaster setting in terms of
maximizing the probability of selecting the most accurate
forecaster. But what if there is more than just a single event?  Let
$\Delta := \min_{j \neq i} \bigl(\max_i R(\bm{y}_i,\theta) -
R(\bm{y}_j,\theta) \bigr)$ denote the difference between the expected
scores of the most accurate forecaster and the second-most accurate
forecaster. Ideally, one would like to guarantee that for any
``accuracy gap'' $\Delta$ and any probability $\pi$ arbitrarily close
to 1, there exists some minimal number of events after which it is
guaranteed that the forecasting competition mechanism selects the most
accurate forecaster with probability at least $\pi$.
This intuition is formally captured in the definition of \emph{limit accuracy}.

\begin{definition}\label{def:limit-accuracy}
  Forecasting competition mechanism $M$ is \emph{limit accurate} with
  respect to proper scoring rule $R$ and set of joint distributions
  $\Theta$ if and only if, for any $n$, any $\underline{\Delta} > 0$,
  and any $\pi \in [0,1)$, there exists a finite number of events
  $\underline{m} \in \mathbb{N}$ such that for all joint distributions
  $\theta \in \Theta$ and all
  $\bm{y}_1, \dots, \bm{y}_n \in [0,1]^m$ with $m \geq \underline{m}$
  and $\Delta > \underline{\Delta}$, it holds that
  $\Pr\limits_{\bm{X} \sim \theta}\Bigl(M\bigl(\bm{y}_1, \dots,
  \bm{y}_n,\bm{X}\bigr) = i\Bigr) > \pi.$
\end{definition}

Proposition~\ref{thm:no-truthful-and-limit-accurate-mechanisms} shows
that some restriction on $\theta$ is necessary as limit accuracy
cannot be achieved for all joint distributions. In particular,
consider the extreme case in which events are ``identical copies'' of
one another, such that whenever $X_1 = 1$, it holds that $X_k = 1$ for all
$k \in \{2, \dots, m\}$ and whenever $X_1 = 0$, we have $X_k = 0$
with $k \in \{2, \dots, m\}$. In that case, all information contained
in events $2, \dots, m$ is already contained in the first event, and
so increasing $m$ is not helpful for identifying the most accurate
forecaster.

\begin{proposition}\label{thm:no-truthful-and-limit-accurate-mechanisms}
  No forecasting competition mechanism is limit accurate for all
  distributions $\theta$ over $X_1, \dots, X_m$.
\end{proposition}

In the remainder of this section, we design a forecasting competition
mechanism that is limit accurate when the events are independent and
strictly incentive compatible when this independence is also reflected
in the uncertainty about others' reports. The restriction on
forecasters' beliefs is referred to as \emph{belief independence}.

\begin{definition}\label{def:belief-independence}
  For joint distribution $D$ over outcomes $\bm{X}$ and reports
  $\bm{Y}_{-i}$, let $D_k$ be the corresponding joint distribution
  over outcome $X_k$ and reports $\bm{Y}_{-i, k}$. $D$ is \emph{belief
    independent} if and only if all $D_k$ for $k \in [m]$ are
  independent.
\end{definition}

Note that under belief independence forecaster $i$ can still believe
that other forecasters are more accurate than herself and also that
others' reports are more accurate on some events than others.

\begin{definition}\label{def:truthful-under-belief-independence}
  Forecasting competition mechanism $M$ is \emph{strictly incentive
    compatible under belief independence} if and only if for all
  forecasters $i \in [n]$, all belief vectors $\bm{p}_i$, all belief
  independent joint distributions $D$ over outcomes $\bm{X}$ and
  reports $\bm{Y}_{-i}$ such that
  $\mathbf{E}_{\bm{X} \sim D}\bigl[\bm{X}\bigr] = \bm{p}_i$, and all
  alternative report vectors $\bm{y'}_i \neq \bm{p}_i$,
  \begin{equation*}
  \Pr\limits_{\bm{X}, \bm{Y}_{-i} \sim D}\bigl(M(\bm{Y}_1, \dots,
  \bm{p}_i, \dots,
  \bm{Y}_n,\bm{X}) = i\bigr) > \Pr\limits_{\bm{X}, \bm{Y}_{-i} \sim D}\bigl(M(\bm{Y}_1, \dots,
  \bm{y'}_i, \dots,
  \bm{Y}_n,\bm{X}) = i\bigr).
\end{equation*}
\end{definition}

\subsection{Incentive-Compatible and Limit-Accurate Mechanism for Independent Events}\label{sec:ielf}

The Independent-Event Lotteries Forecasting Competition Mechanism
(\ielf) $\MIELFR(\bm{y}_1, \dots, \bm{y}_n,\bm{x})$ is defined as:
\begin{enumerate}
\item For each event $k$, pick forecaster $i$ to be the event winner
  $w_k \in [n]$ with probability
  \begin{equation*}
    f_{i,k}(y_{1,k}, \dots, y_{n,k},x_k) = \frac{1}{n} + \frac{1}{n} \Bigl(R\bigl(\yik,x_k\bigr)-\frac{1}{n-1} \sum_{j
      \neq i} R\bigl(y_{j,k},x_k\bigr)\Bigr).
  \end{equation*}
  where $R \in [0,1]$ if $m=1$ and $R \in [0,1)$ if $m \geq 2$ is a
  bounded strictly proper scoring rule.\footnote{If used in
    conjunction with a normalized $R$ for $m \geq 2$, $\MIELFR$ may
    fail to be strictly incentive compatible (it is still weakly
    incentive compatible) when there exists an event for which a
    forecaster believes that she is a perfect forecaster reporting
    100\% for the eventually materialized outcome and every other
    forecaster is doing the opposite, i.e., reporting 0\% for the
    eventually materialized outcome. We do not expect this to be an
    issue in practical application.}

\item Select the forecaster who won the most events, $\argmax_i
  \sum_{k=1}^m \mathbbm{1}(w_k = i)$, breaking ties uniformly at
  random. Here $\mathbbm{1}$ denotes the 0/1 indicator function.
\end{enumerate}

In essence, \ielf runs a single \elf lottery for each event and awards
the prize to the forecaster who won the most lotteries.

\begin{theorem} \label{thm:multi-event-truthful} $\MIELFR$ is strictly
  incentive compatible under belief independence for $m \ge 1$ events.
\end{theorem}

Take the perspective of any forecaster $i \in [n]$ seeking to
maximize the probability of being selected. 
The proof proceeds by
showing that she can reason about each event independently
because of belief independence and, in a second step, that
increasing her probability of winning event $k$ strictly increases her
probability of winning overall.

To conclude this section, we show that \ielf is limit accurate when events are independent.

\begin{theorem}\label{thm:multi-event-accuracy}
  $\MIELFR$ is limit accurate for all $R$ and all $\theta$ such that
  event outcomes $X_1, \dots, X_m$ are independent.
\end{theorem}

For intuition, note that more accurate forecasters have a higher
probability of winning each event (by
Proposition~\ref{prop:single-event-accuracy}). Hence, by standard
concentration inequality arguments, the most accurate forecaster wins
the most events with high probability when events are independent and
the number of events is large.

Note that, as an alternative to \ielf, one could collapse the $m$
binary random variables into a single categorical random variable with
$2^m$ outcomes and apply \elf to the joint distribution implied by the
forecasters' (marginal) reports. (As we discuss in
Section~\ref{sec:categorical}, \elf readily extends to the categorical
case.) The problem with this mechanism is that it is not limit
accurate. In particular, it will not select the most accurate
forecaster with probability higher than $2/n$. To see this, observe
that, in Equation~\ref{eq:se-elf}, the first term in the parentheses
is at most 1 and the second term at least 0, resulting in at most
$2/n$.

\section{Discussion}
\label{sec:discussion}

In this section, we describe extensions to our model and
discuss the practical implementation of our methods.

\subsection{Categorical Outcomes}
\label{sec:categorical}

So far, we have restricted our analysis to events with binary
outcomes. In practice, we are also interested in events with
non-binary (categorical) outcomes.
Unsurprisingly, selecting the forecaster with highest average proper
score (e.g., using Brier's~\shortcite{brier:1950} categorical
generalization of the quadratic scoring rule) inherits the violation
of incentive compatibility exhibited in Section~\ref{sec:psr}.

\elf readily extends to categorical outcomes.
The competitive scoring rule of \cite{kilgour-gerchak:2004} is
incentive compatible for categorical outcomes when used in conjunction
with any proper multi-outcome scoring rule, and \elf inherits this
incentive compatibility for all such rules that are bounded. Under
belief independence, incentive compatibility of \ielf follows from the
same arguments used in the proof of
Theorem~\ref{thm:multi-event-truthful}. Moreover, it still holds that
more accurate forecasters obtain higher scores in expectation, so the
most accurate forecaster still wins the most events in
expectation. Hence, we can prove limit accuracy by a qualitatively
identical argument to the one in the proof of
Theorem~\ref{thm:multi-event-accuracy}.

\subsection{Real-Valued Outcomes and Reports}

In many business contexts, we are interested in forecasting events
that take real-valued outcomes instead of categorical values. For
instance, events could be the monthly demand of particular items, the
cost of infrastructure projects, or the annual inflation rate.
 Both \elf and \ielf readily extend to handle these
cases. In contrast to events with categorical outcomes,
where one typically seeks to elicit the forecaster's entire subjective
probability distribution over the outcomes, this is cumbersome with
infinitely many outcomes on the real line. Instead, practitioners
typically choose to only elicit properties of the underlying
distribution, such as the mean or the median, which summarize the
underlying distribution in ways meaningful for the application at
hand. There exist many proper scoring rules for the elicitation of
these properties.
For example, it is well known that the quadratic scoring rule
$R_q(y,x)=1-(y-x)^2$, which was introduced in Section~\ref{sec:psr},
generalizes to real valued outcomes $x \in [0,1]$. More precisely, if
random variable $X$ denotes the real-valued outcome, the forecaster
maximizes her expected score by reporting $y = \E[X]$, i.e., her
subjective estimate of the mean of $X$. Meanwhile, the absolute
scoring rule $R_a(y,x)=1-|y-x|$ is strictly proper when used to elicit
subjective estimates of the median of $X$ \egcitep{jose:2017}. Note
that these scoring rules can be scaled to incorporate any bounded
interval $[a,b]$ with $b > a$. Moreover, while it is easy to obtain
upper and lower bounds on the variable of interest for almost any
conceivable application, tighter bounds yield better discrimination in
score between more and less accurate reports.

While the quadratic and absolute scoring rules are strictly proper
when used as payments to elicit subjective estimates of the mean and
median, respectively, misreporting remains an issue when they are
naively applied to forecasting competitions. Consider random variable
$X$ commonly known to be uniformly distributed on $[0,1]$. If $n=3$
forecasters all report a subjective estimate of the mean, i.e.,
$y_i=0.5$ for all $i$, and the forecaster with highest quadratic score
is selected as the prize winner, then each forecaster wins with
probability $1/3$ (assuming ties are broken uniformly at
random). However, if forecaster $1$ instead reports $y_1=0.5-\epsilon$
for some small $\epsilon$, then she achieves the highest score
whenever $X<0.5-\epsilon$, which occurs with probability
$0.5-\epsilon > 1/3$.\footnote{Observe that this misreport is somewhat
  different from those in the categorical setting, where rational
  forecasters will generally ``extremize'' their reports towards an
  outcome. In contrast, in the example above, a forecaster who
  unilaterally deviates to reporting an extreme value of 0 or 1 would
  only be selected with probability $1/4$.} The same example continues
to break incentive compatibility when the absolute scoring rule is
used to elicit estimates of the median.
  
To overcome the issue of misreporting, we can define \elf and \ielf as
in Sections~\ref{sec:single-event} and~\ref{sec:accuracy}, just with
an appropriately chosen scoring rule $R$ that is strictly proper for
the property being elicited. Strict incentive compatibility of \elf
and \ielf (under the belief independence restriction) follows by
reasoning analogous to the binary case. The accuracy guarantee
provided by \ielf carries over as well, with the accuracy implied by
the scoring rule $R$ used to define the mechanism. As for the
binary-outcome setting, both \elf and \ielf work in conjunction with
any bounded $R$. Observe that this is analogous to using proper
scoring rules as payments, where $R$ needs to be bounded to guarantee
non-negative payments.

\subsection{Outputting a Forecaster Ranking}

In some practical applications, it may be more appropriate to output a
ranking rather than a single forecaster. For example, most play-money
prediction markets maintain a ranking of contestants. Similarly, many
Kaggle competitions award prizes to the highest-ranked forecasters
with prizes decreasing in value as the forecasters' ranks increase.
Ranking forecasters in order of any proper score again
inherits all of the problems described in Section~\ref{sec:psr}.

\ielf can
be adapted to produce a ranking by simply ordering forecasters
according to the number of events that the forecasters win.
As long as forecasters strictly prefer higher positions in the ranking
(e.g., because higher rankings correspond to higher-valued prizes),
\ielf remains strictly incentive compatible, since forecasters
maximize their probability of winning an event (and potentially moving
up in the ranking) by reporting truthfully.
Moreover, the same style of accuracy results from
Section~\ref{sec:ielf} hold, at least
qualitatively, when the objective is to maximize the probability of
outputting the correct ranking. In expectation, more accurate
forecasters achieve higher proper scores, leading to higher expected
values of $f_{i,k}$. Thus, more accurate forecasters win more events
in the long run, and the true ranking is faithfully revealed.

\subsection{Forecaster Hiring and Connections to Learning}
\label{sec:pac-learning}

Forecasting competitions are often used as a method of selecting a
forecaster to hire when future predictions are needed. In this
setting, the goal of the competition mechanism is to select the
forecaster who will be (approximately) the most accurate on future
events. There is an implicit assumption here that good performance on
the observed events translates into good performance in the future, a
well-established fact in practice \egcitep{mellers-et-al:2014}.

Our methods and results can be extended to this setting. Instead of
determining accuracy through the $m$ events being predicted, we could
instead assume a joint distribution $D_{\theta}$ over event
probabilities $\theta$ and the beliefs $p_i$ of each forecaster
$i$. We could then define the accuracy of forecaster $i$ in terms of
the expected proper score of her truthful forecasts with respect to
$D_{\theta}$, i.e.,
$\E_{p_i, \theta \sim D_{\theta}}\left[R(p_i, \theta)\right]$.

Under this model, mechanism $\MPSR$ discussed in Section~\ref{sec:psr}
can be viewed as performing an analog of empirical risk
minimization. Similar to how basic empirical risk minimization bounds
are proved for PAC learning~\citep{kearns:1994}, we could then argue
that, with high probability, the forecaster with the highest score on
any observed sample of events has expected accuracy close to that of
the best forecaster in the set. Therefore, as the number of events
grows large, the forecaster selected by $\MPSR$ would be guaranteed to
have accuracy arbitrarily close to that of the most accurate
forecaster. However, the incentive issues remain. The advantage of
\ielf is that it obtains truthful reports for any $m$ while achieving
similar accuracy guarantees as $m$ grows large. In this sense, \ielf
can be viewed as a mechanism for learning in the presence of strategic
agents, where the objective is to select a forecaster that will
perform well on future events.

\subsection{Practical Implementation}\label{sec:practical-implementation}

In Section~\ref{sec:model}, we require that all forecasters report
their predictions for all events before the first event
materializes. With an appropriate generalization of the definition of
incentive compatibility, this requirement can be relaxed without
sacrificing the properties of \elf. In particular, when reporting on
event $k$, we can allow forecasters to update joint distribution $D$
conditioned on the outcomes of past events and the reports on these
events. Our results continue to hold if incentive compatibility
requires that forecasters truthfully report their updated beliefs.

For \ielf, suppose that a forecaster reports on event $k$ after some
subset of the other events have materialized.
Given belief independence, the reports of other forecasters on any
other event, as well as the corresponding outcomes for any events
already materialized, do not lead to a belief update.
Therefore, the competition organizer does not need to protect or
withhold any information from the forecasters as long as the
randomness involved in selecting event winners $w_k$ from
probabilities $f_{i,k}$ is not realized until all predictions have
been reported.

More speculatively, one could imagine applying our
  techniques to other elicitation methods. In particular, prediction
  markets are often implemented using play money, with monetary prizes
  for top-ranked traders or simply high positions on public
  leaderboards used as incentive~\egcitep{jia-et-al:2017}. Directly
  awarding prizes to participants with the highest play money account
  balances, however, leads to incentive problems analogous to those in
  the forecasting competition model we consider in this paper:
  maximizing the probability of having the highest account balance is
  not the same as maximizing expected account balance. Variations on
  this idea induce similar gaming incentives; for example,
  \citet{chakraborty:2013} award prizes uniformly at random among
  participants placed sufficiently high on the leaderboard. While it
  is not clear how to directly translate (I-)ELF to this setting, it
  is easy to see that awarding a single prize randomly with
  probability proportional to account balance does lead to forecasters
  maximizing their expected account balance\footnote{This assumes that
    no money leaves the system in the form of fees or
    withdrawals, a reasonable assumption for play money
    markets.}~\citep[e.g.,][Section~1.2.2]{cowgill-zitzewitz:2015}.
  Further exploring applications to prediction markets and other
  elicitation methods is a compelling direction for future work.

Note that both \elf and \ielf are easy to implement. Indeed,
even for very large competitions, both mechanisms can be implemented
in standard spreadsheet software. Each value $f_{i,k}$ is computed by
a simple formula, after which the only remaining step is to implement
1 or $m$ lotteries for \elf and \ielf, respectively.

\section{Conclusion}
\label{sec:conclusion}

In real-world forecasting settings, forecasters typically compete for
a single prize. Motivated by the prevalence of these forecasting
competitions and their poor incentive properties, we initiate the
study of incentive-compatible forecasting competitions.
Despite a rich literature on incentive-compatible forecast elicitation
in the non-competitive setting, the mechanisms in this work are the
first to solve the incentive challenge in the competition setting.
The forecasting competition mechanism most widely used in practice is
to simply select the forecaster with highest score according to some
proper scoring rule. Not only does this particular mechanism fail to
elicit truthful forecasts, but, as we show, any deterministic
forecasting competition mechanism must violate incentive
compatibility.
We therefore turn to randomized forecasting competitions, which can be
thought of as rewarding forecasters with a lottery ticket that has a
higher chance of winning the more accurate the forecaster was relative
to the other forecasters in the competition. This intuitive principle
is behind both mechanisms we design.

We first define the Event Lotteries Forecasting Competition Mechanism
(\elf), which incentivizes truthful reports for arbitrary beliefs on
behalf of the forecasters. Due to its randomized nature, \elf may not
always select the most accurate forecaster, but it does select more
accurate forecasters with higher probability than less accurate
ones. For the special case of one event and two forecasters, we show
that, under mild technical conditions, no incentive-compatible
mechanism can select the most accurate forecaster with higher
probability than \elf does.

Our second mechanism, \ielf, is strictly incentive compatible when
forecasters' beliefs satisfy belief independence, which, intuitively,
requires that information about one event does not inform forecasters'
beliefs about other events. \ielf uses \elf as a building block, first
selecting a winner for each event using \elf, and then selecting the
competition winner as the forecaster who won the most individual
events. In addition to being incentive compatible under belief
independence, \ielf also selects the most accurate forecaster with a
probability that tends to 1 as the number of events grows.

Our results have significant implications for organizations that
employ groups of forecasters to inform managerial decision making
under uncertainty. Previous studies on forecasters' competitive
incentives encouraged the fostering of collaboration and cooperation
to mitigate the distorted incentives at
play~\citep{lichtendahl-winkler:2007}. Our work yields a different
perspective. By cleverly exploiting randomization, the decision maker
can embrace competitive stakes when eliciting predictions without
having to sacrifice the quality of the information received.

\begin{APPENDIX}{}

\section{Generalizing Immutable-Belief Incentive Compatibility
  to Robust Incentive Compatibility}\label{sec:bayesian-model}

In this section, we are going to unpack how the (standard)
immutable-belief model---while appropriate for the wagering setting,
where different forecasters, by definition, agree to disagree and seek
to bet on their individual
convictions~\citep{lambert-et-al:2008a}---is too limited for
forecasting settings, where forecasters believe that other
forecasters' reports contain information that they themselves do not
already have. This includes but is not limited to the competition
setting that is the focus of this paper. The section is organized as
follows. First, we provide the definition of immutable-belief
incentive compatibility due to Kilgour and
Gerchak~\shortcite{kilgour-gerchak:2004} and Lambert et
al.~\shortcite{lambert-et-al:2008a} and show that it is a special case
of the definition used in the main text of this paper. Second, we
provide an example of a Bayesian belief model along the lines of
standard models in the literature, which is incompatible with the
assumption of immutable beliefs since forecasters update their beliefs
about the outcome when learning the beliefs of other forecasters.
Finally, using a particular numerical example, we demonstrate that an
immutable-belief incentive compatible mechanism that has been
suggested in the literature incentivizes misreports under this
Bayesian model.

We emphasize here that neither ELF nor I-ELF assume that beliefs are
formed using this particular model. We also emphasize that our
robust incentive compatibility generalizes both Bayesian and
immutable-belief models. In particular, truthful reporting is a
dominant strategy in both ELF and I-ELF, regardless of whether agents
would update their beliefs knowing the reports of other agents or
not. Hence, this Bayesian model is given here for illustrative
purposes only, showing that mechanisms that are incentive compatible
only for immutable beliefs are not sufficient when forecasters believe
that other forecasters' reports contain information that they
themselves do not already have.
We now state the incentive compatibility definition, applied to the
competition setting, that was used in the work of Kilgour and
Gerchak~\shortcite{kilgour-gerchak:2004} and Lambert et
al.~\shortcite{lambert-et-al:2008a}.
\begin{definition}\label{def:truthful-immutable}
  Forecasting competition mechanism $M$ is \emph{strictly incentive
    compatible for immutable beliefs} if and only if for all
  forecasters $i \in [n]$, all belief vectors $\bm{p}_i$, all others'
  reports $\bm{y}_{-i}$, and all alternative report vectors
  $\bm{y'}_i \neq \bm{p}_i$,
  $\Pr\limits_{\bm{X} \sim p_i}\bigl(M(\bm{y}_1, \dots, \bm{p}_i,
  \dots, \bm{y}_n,\bm{X}) = i\bigr) > \Pr\limits_{\bm{X} \sim
    p_i}\bigl(M(\bm{y}_1, \dots, \bm{y'}_i, \dots, \bm{y}_n,\bm{X}) =
  i\bigr)$.
\end{definition}

Observe that Definition~\ref{def:truthful-immutable} coincides with
the (robust) incentive compatibility definition used in this work
(Definition~\ref{def:truthful}) when joint distribution $D$ is
restricted such that $\bm{y}_{-i}$ only takes a single value,
regardless of the realization of $X$.
Hence, every mechanism that is robust incentive compatible
(Definition~\ref{def:truthful}) is also incentive compatible for
immutable beliefs (Definition~\ref{def:truthful-immutable}). However,
the reverse is not true.
For intuition as to how robust incentive compatibility is different
from immutable-belief incentive compatibility and to understand why
one wants forecasting competition mechanisms to satisfy the stronger
robust incentive compatibility, ignore for a moment that, in
competitions, ``payments'' (selection probabilities) need to add up to
1. Consider then a forecaster $i$ who is paid $y_j \cdot R_q(y_i,x)$,
where $y_j$ is the report of another forecaster $j \neq i$. In the
immutable-belief model, $y_j$ is assumed to be a constant from
forecaster $i$'s perspective, so that she should report truthfully
because linear transformations of proper scoring rules preserve
properness. However, if forecaster $i$ believes $j$'s report to be
correlated with outcome $X$, then $j$'s report $Y_j$ is in fact a
random variable and not a constant. This typically leads to
misreports. In the extreme case, if forecaster $i$ believes that
forecaster $j$ reports all probability mass on the eventually
materialized outcome, i.e., $Y_j = X$, then, if $X=0$, she receives
payment $0$, and if $X=1$, she receives $R_q(y_i,1)$. Thus, forecaster
$i$ strategizes by conditioning on $X=1$, maximizing her payment by
reporting $y_i = 1$ regardless of her true belief. As we will see
later in this section, this intuition also applies to competition
settings, including settings where forecaster $i$ believes that she is
more accurate than all other forecasters.

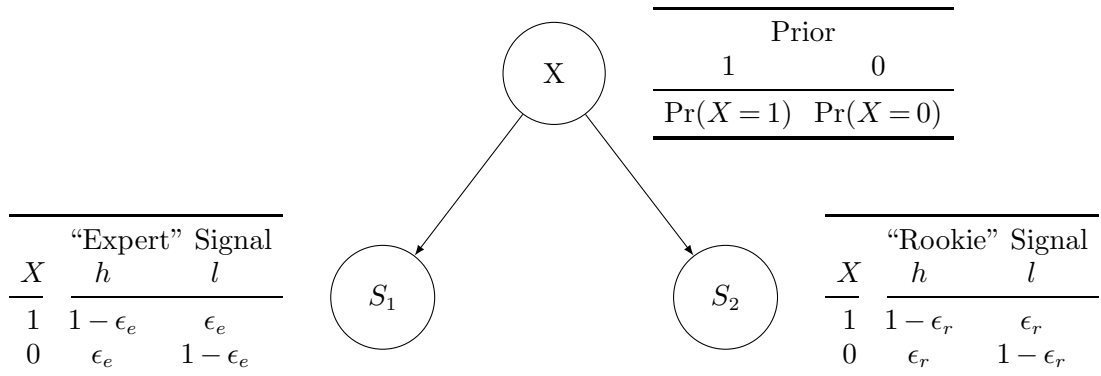
\begin{figure}[ht]
\begin{tikzpicture}[
  node distance=2cm and 1.3cm,
  mynode/.style={draw,circle,text width=1cm,align=center}
]
\node[mynode] (sp) {X};
\node[mynode,below left=of sp] (gw) {$S_1$};
\node[mynode,below right=of sp] (ra) {$S_2$};
\path (sp) edge[-latex] (ra)
(sp) edge[-latex] (gw);
\node[left=0.5cm of gw]
{
\begin{tabular}{ccc}
\toprule
& \multicolumn{2}{c}{``Expert'' Signal} \\
$X$ & \multicolumn{1}{c}{$h$} & \multicolumn{1}{c}{$l$} \\
\cmidrule(r){1-1}\cmidrule(l){2-3}
1 & $1-\epsilon_e$ & $\epsilon_e$ \\
0 & $\epsilon_e$ & $1-\epsilon_e$ \\
\bottomrule
\end{tabular}
};
\node[right=0.5cm of sp]
{
\begin{tabular}{cc}
\toprule
\multicolumn{2}{c}{Prior} \\
\multicolumn{1}{c}{1} & \multicolumn{1}{c}{0} \\
\cmidrule{1-2}
$\Pr(X=1)$ & $\Pr(X=0)$ \\
\bottomrule
\end{tabular}
};
\node[right=0.5cm of ra]
{
\begin{tabular}{ccc}
\toprule
& \multicolumn{2}{c}{``Rookie'' Signal} \\
$X$ & \multicolumn{1}{c}{$h$} & \multicolumn{1}{c}{$l$} \\
\cmidrule(r){1-1}\cmidrule(l){2-3}
$1$ & $1-\epsilon_r$ & $\epsilon_r$ \\
$0$ & $\epsilon_r$ & $1-\epsilon_r$ \\
\bottomrule
\end{tabular}
};

\end{tikzpicture}
\caption{Example of a Bayesian forecaster model with $n=2$. Forecaster
  $1$ is of the ``expert'' type, forecaster $2$ is of the ``rookie''
  type.\label{fig:model}}
\end{figure}

In contrast to immutable-belief incentive compatibility, robust
incentive compatibility does guarantee truthful reporting incentives
even in settings in which forecaster $i$ would update her belief upon
learning forecaster $j$'s report.
Note that such conditional belief updating is implied by standard
Bayesian models in the forecasting literature where individual
forecasters' beliefs stem from noisy observations of some ground
truth~\egcitep{lichtendahl-winkler:2007,lichtendahl-et-al:2013,palley-soll:2019}.
To make this concrete, consider the following simple Bayesian model
along those lines. (An example of this model is depicted in
Figure~\ref{fig:model}; multiple-event models can be defined
analogously.) The event outcome is given by random variable $X$, which
takes values in $\{0,1\}$. All $n$ forecasters share a common prior
$\Pr(X=1)$ that the event outcome is $1$ (e.g., a commonly-known base
rate). Before the event outcome materializes, each forecaster
$i \in [n]$ observes a binary, noisy signal $S_i$, taking values in
$\{l,h\}$. Each forecaster is of one of two types: ``expert'' types
have a noise level (error rate) of $\epsilon_e$ and forecasters of
``rookie'' types have a noise level of $\epsilon_r$ with
$0.5 > \epsilon_r > \epsilon_e > 0$. If $X=1$, the probability of
observing $h$ is $1-\epsilon$, and if $X=0$, the probability of
observing $l$ is $1-\epsilon$, where the $\epsilon$ value depends on
the forecaster type. The belief model as well as which forecaster is
of which type is common knowledge.  After observing her signal
$S_i = s_i$, forecaster $i$ updates her belief about $X$. Moreover,
she also updates her (meta) beliefs about the beliefs of the other
forecasters conditional on that $X$. The updated belief on the outcome
is given by
\begin{equation}\label{eq:bayes}
  \Pr(X = 1| S = s) = \frac{\Pr(S=s|X=1) \cdot \Pr(X=1)}{\Pr(S=s)},
\end{equation}
where $\Pr(S=s|X=1)$ depends on the forecaster's noise level as
determined by her type and
\begin{equation}
  \Pr(S = s) = \Pr(S=s|X=1) \cdot \Pr(X=1) + \Pr(S=s|X=0) \cdot \Pr(X=0).
\end{equation}
For the meta beliefs of forecaster $i$ about the belief of forecaster
$j$ given $X$, first observe that forecaster $j$ can only hold one of
two possible beliefs, namely $\Pr(X=1|S_j)$ for $S_j = h$ and
$S_j = l$, respectively. Which of these two beliefs forecaster $j$
holds is thus determined by her signal, which itself
is influenced by $X$.
In particular, the expected value of forecaster $j$'s belief given
each possible instantiation of $X$ is calculated by
\begin{equation}\label{eq:meta}
  \E\bigl[P_j | X = x\bigr] = \Pr(X=1|S_j = h) \cdot \Pr(S_j = h | X=x) + \Pr(X=1|S_j = l) \cdot \Pr(S_j = l | X=x),
\end{equation}
where $P_j$ denotes the random variable for forecaster $j$'s
belief $p_j$.

To see that immutable-belief incentive compatibility
(Definition~\ref{def:truthful-immutable}) is inappropriate for this
kind of Bayesian model from a technical perspective, it is sufficient
to observe that truthful reports of the other forecasters (i.e.,
beliefs) do indeed depend on the realization of $X$. In particular,
the forecasters' beliefs are correlated with the outcome.
Unfortunately, this observation is not just a technical nuisance but
has immediate implications on mechanisms suggested in the
literature. In the remainder of this section, we will show that
immutable-belief incentive compatible mechanisms suggested in the
literature do indeed lead to misreports in Bayesian models such as the
one exemplified here.

More precisely, we consider a member of the \emph{adaptive weighted
  score} mechanism family suggested in Section 6.1
of~\cite{lambert-et-al:2008a}, which we present here applied to the
forecasting competition setting. This family of mechanisms is
parameterized by a choice of scoring rule $R$. We note here that the
particular choice we make is not an edge case. For simplicity, we
present the example mechanism for $n=4$. Intuitively, the mechanism
repeatedly partitions the 4 forecasters into two groups $A$ and
$\bar{A}$ of 2 forecasters each, and scores the forecasters in the
first group ``against'' each other using a scheme similar to that of
Kilgour and Gerchak~\shortcite{kilgour-gerchak:2004}. The interesting
part is that the proper scoring rule that is used to score forecasters
in the first group is defined by the reports of the second group. As
we will see, this mechanism is immutable-belief incentive compatible
but leads to misreports in the Bayesian model we just introduced.

The mechanism proceeds as follows:
\begin{enumerate}
\item Given the set of $4$ players $\{1, 2, 3, 4\}$, we consider the
  set of forecaster groups of size $2$, which we denote by
  $\mathcal{A}$. Further, let $\mathcal{A}_i \subset \mathcal{A}$
  denote the set of forecaster groups that contain forecaster $i$.
\item Let
  $R^{z, z'}(y,x) = \frac{z+z'+\varepsilon}{2+\varepsilon} \cdot
  R_q(y,x)$ with $\varepsilon > 0$ be a strictly proper scoring rule,
  whose form is parameterized by $z, z' \in [0,1]$. Note that for any
  constants $z, z'$, $R^{z, z'}$ is a (weakly) scaled-down version of
  the (normalized) quadratic scoring rule $R_q$. Further note that the
  role of $\varepsilon$ is to ensure that the scaling factor is
  positive even if $z, z' = 0$, so that $R^{z, z'}$ remains strictly
  proper in that case. Observe that $R^{z, z'}$ is bounded between 0
  and 1.
\item For a single event and $n=4$, the {\em Adaptive-Score
    Forecasting Competition Mechanism} $\MASFR(y_1, \dots, y_n,x)$
  selects forecaster $i \in [n]$ with probability
  \begin{equation*}
    f_i(y_1, \dots, y_n,x) = \frac{1}{4} + \sum_{A \in \mathcal{A}_i}\frac{1}{12} \biggl(R^{z, z'}\bigl(y_i,x\bigr)-R^{z, z'}\bigl(y_j,x\bigr)\biggr),
  \end{equation*}
  where $j\in A$ refers to the other forecaster $j \neq i$ 
  in each $A \in \mathcal{A}_i$, and $z, z'$ are the two reports from
  forecaster group $\bar{A}$ not containing $i$, i.e.,
  $\bar{A} := [n] \backslash A$.
\end{enumerate}

It is easy to see that $\MASFR$ is strictly incentive compatible for
immutable beliefs (also see Lambert et
al.~\shortcite{lambert-et-al:2008a}): if forecaster $i$ believes that
the other forecasters' reports $y_j, z, z'$ are constants, which are
uninformative about $X$, then, for each $A \in \mathcal{A}_i$,
forecaster $i$ believes that she is scored by a scaled-down $R_q$, and
hence should report truthfully. Alternatively, one can think of the
immutable-belief setup as though the reports of all forecasters are
known beforehand, which is explicit in a wagering setting, where, by
definition, forecasters agree to disagree. That is, there is no
uncertainty about the reports of the other forecasters and hence also
no uncertainty about the scoring rule that will be used. The only
uncertainty that remains is about the outcome.

In the remainder of this section, we will show that despite $\MASFR$
being strictly incentive compatible for immutable beliefs, forecasters
can have incentives to misreport in the Bayesian model described
earlier in this section. It is important to emphasize that none of
this is an edge case: other families of immutable-belief incentive
compatible mechanisms, other choices of scoring rules for this family,
and other numbers for this particular choice of scoring rule would
also lead to misreporting incentives in a Bayesian context.

The numerical example setting we consider has $n=4$ forecasters and a
uniform prior of $\Pr(X=1) = \Pr(X=0) = 0.5$. Further, forecaster
$1$ is of
the expert type with $\epsilon_e = 0.2$, and forecasters $2$, $3$, and
$4$ are of the rookie type with $\epsilon_r = 0.3$. For scoring rule
$R^{z, z'}$, we use $\varepsilon = 0.1$.

We now consider the situation of forecaster $1$ and show that she has
an incentive to misreport in the special case of all other forecasters
reporting truthfully, i.e., $y_j = p_j$ for all $j \neq 1$. (Remember
that both definitions of incentive compatibility are with respect to
dominant strategies, which require that truthful reporting is
maximizing forecaster $1$'s selection probability for \emph{any}
reports of the other forecasters.) Forecaster $1$'s (expected)
selection probability is
\begin{equation}\label{eq:exp-selection-prob}
  \begin{split}
    \E\bigl[f_i(Y_1, \dots, y_i, \dots, Y_n,X)\bigr] =\, &\frac{1}{4}
    + \sum_{A \in \mathcal{A}_i}\frac{1}{12} \E\biggl[R^{Z,
      Z'}\bigl(y_i,X\bigr)-R^{Z, Z'}\bigl(Y_j,X\bigr)\biggr],
  \end{split}
\end{equation}
where the expectation is taken over the randomness of the Bayesian
model. $\mathcal{A}_1$ contains forecaster groups $\{1,2\}$,
$\{1,3\}$, and $\{1,4\}$. Since forecasters $2$, $3$, and $4$ are of
the rookie type, forecaster $j$ from $A \in \mathcal{A}_1$ is always a
rookie and reports $z$ and $z'$ from $\bar{A}$ are also from
rookies. Hence, forecaster $i$'s expected score for forecaster groups
$\{1,2\}$, $\{1,3\}$, and $\{1,4\}$ are the same, and so we first
consider only forecaster group $A = \{1,2\}$ and later multiply the
expected score for that group by $3$.

Rookie types have one of two possible beliefs about the outcome,
depending on which signal they observed (Eq.~\ref{eq:bayes}). For
forecaster $2$, this is either
$\Pr(X=1|S_2 = h) = \frac{\Pr(S_2=h|X=1) \cdot \Pr(X=1)}{\Pr(S_2=h)} =
\frac{0.7 \cdot 0.5}{0.5} = 0.7$ or $\Pr(X=1|S_2 = l) = 0.3$.  Since
$\bar{A} = \{3,4\}$ also contains only rookies, their possible beliefs
are the same as for forecaster $2$. Thus, scoring rule $R^{z, z'}$ has
three possible scaling factors
$\frac{z+z'+\varepsilon}{2+\varepsilon}$ for $R_q$, which depend on
the reports of the forecasters in $\bar{A} = \{3,4\}$, namely
$\frac{0.7+0.7+0.1}{2+0.1} = \frac{5}{7}$,
$\frac{0.7+0.3+0.1}{2+0.1} = \frac{11}{21}$, and
$\frac{0.3+0.3+0.1}{2+0.1} = \frac{1}{3}$.
Using notation $S_{\bar{A}}$ to denote the signals observed by the
forecasters in $\bar{A}$, the probabilities for the first and second
scaling given $X=x$ can, due to conditional independence of $S_3$ and
$S_4$, be calculated by (the third is calculated analogously to the
first)
\begin{equation*}
  \Pr(S_{\bar{A}} = \{h,h\}| X = x) = \Pr(S_3 = h|X=x) \cdot \Pr(S_4 = h|X=x)
\end{equation*}
and
\begin{equation*}
  \Pr(S_{\bar{A}} = \{l,h\}| X = x) = \Pr(S_3 = h|X=x) \cdot \Pr(S_4 = l|X=x) + \Pr(S_3 = l|X=x) \cdot \Pr(S_4 = h|X=x).
\end{equation*}
This results in $\Pr(S_{\bar{A}} = \{h,h\}| X = 1) = \Pr(S_{\bar{A}} = \{l,l\}| X = 0) = 0.49$, $\Pr(S_{\bar{A}} = \{l,h\}| X = 1) = \Pr(S_{\bar{A}} = \{l,h\}| X = 0) = 0.42$, and $\Pr(S_{\bar{A}} = \{l,l\}| X = 1) = \Pr(S_{\bar{A}} = \{h,h\}| X = 0) = 0.09$.
With this, forecaster $1$ can now reason about the scoring rule she
expects for each event outcome. If $X=1$, forecaster $1$ expects to be
scored by scoring rule
$\left(0.49 \cdot \frac{5}{7} + 0.42 \cdot \frac{11}{21} + 0.09 \cdot
  \frac{1}{3}\right) R_q(y_1,1) = 0.6 \cdot R_q(y_1,1)$. Analogously, if
$X=0$, she expects scoring rule
$\left(0.09 \cdot \frac{5}{7} + 0.42 \cdot \frac{11}{21} + 0.49 \cdot
  \frac{1}{3}\right) R_q(y_1,0) = \frac{47}{105} \cdot R_q(y_1,0) = 0.448
\cdot R_q(y_1,0)$.

Forecaster $1$'s belief about forecaster $2$'s report given $X$ is
calculated by Eq.~\ref{eq:meta} and results in
$\E\bigl[Y_2 | X = 1\bigr] = 0.7 \cdot 0.7 + 0.3 \cdot 0.3 = 0.58$ and
$\E\bigl[Y_2 | X = 0\bigr] = 0.7 \cdot 0.3 + 0.3 \cdot 0.7 = 0.42.$ If
$X=1$, the expectation in the right hand side of
Eq.~\ref{eq:exp-selection-prob} for $A = \{1,2\}$ is then
\begin{equation*}
  \E \Bigl[R^{Y_3, Y_4}\bigl(y_1,1\bigr)-R^{Y_3, Y_4}\bigl(Y_2,1\bigr) | X = 1\Bigr] =  0.6 \cdot \Bigl(1-(y_1-1)^2 - \left(1-(0.58-1)^2\right)\Bigr) = 0.106 - 0.6\, (y_1-1)^2, 
\end{equation*}
where the expectation is again taken over the randomness of the
Bayesian model. Analogously, if $X=0$, her expectation for that part
is
\begin{equation*}
\E \Bigl[R^{Y_3, Y_4}\bigl(y_1,0\bigr)-R^{Y_3, Y_4}\bigl(Y_2,0\bigr) | X = 0\Bigr] = 0.448 \cdot \Bigl(1-y_1^2 - \left(1-0.42^2\right)\Bigr) = 0.079 - 0.448\, y_1^2. 
\end{equation*}
Observe that in each outcome, forecaster $1$ is scored using a
positive-affine transformed $R_q$. Crucially however, the scaling
factor is higher for $X=1$ than for $X=0$. As we will see, this has
the effect that forecaster $1$ has an incentive to shift her report
towards the $X=1$ outcome as it carries more weight.
To obtain forecaster $1$'s overall expected scores for each $X$, we
multiply the expected scores for $A = \{1,2\}$ by $3$ (to account for
the symmetric cases of $A = \{1,3\}$ and $A = \{1,4\}$), divide the
result by $12$ (resulting in a division by $4$), and add
$\frac{1}{4}$.

To complete the example, suppose that forecaster $1$ observes
$S_1 = h$. Using Eq.~\ref{eq:bayes}, she updates her belief about the
outcome to $\Pr(X=1|S_1 = h) = 0.8$.
Putting this all together, forecaster $1$'s expected score reporting
$y_1$ is
\begin{equation*}
  \begin{split}
    \E\bigl[f_i(Y_1, \dots, y_i, \dots, Y_n,X)\bigr] =\, &\frac{1}{4}
    + \frac{1}{4} \Bigl( 0.8 \cdot \bigl(0.106 - 0.6\, (y_1-1)^2\bigr) + 0.2 \cdot \bigl(0.079 - 0.448\, y_1^2\bigr)\Bigr),
  \end{split}
\end{equation*}                                                  %
which is uniquely maximized for $y_1 = \frac{75}{89} =
0.843$. Forecaster $1$ thus has an incentive to misreport her true
belief of $0.8$.
It is important to note here that while the exact calculations are
rather extensive, forecasters in this Bayesian setting faced with this
mechanism do not need to compute their conditional beliefs precisely
but can simply make a report that is slightly higher than their belief.

We emphasize that this example also shows that even if a forecaster
believes that she is the most accurate forecaster, she may still have
an incentive to misreport under immutable-belief incentive
compatibility. The key advantage of robust incentive compatibility
over immutable-belief incentive compatibility is that it allows for
the possibility that forecasters may believe that other forecasters'
reports contain some information they do not already have. Or, phrased
differently, in contrast to immutable-belief incentive compatibility,
robust incentive compatibility allows for the possibility that
forecasters would update their beliefs upon learning the reports of
other forecasters.

\section{Procedure to Normalize a Proper Scoring Rule}\label{sec:normalization}

Let $R$ be a bounded proper scoring rule with
$\underline{R} = \min_{y,x} R(y,x)$ and
$\overline{R} = \max_{y,x} R(y,x)$ for
$y \in [0,1] ,\, x \in \{0,1\}$. Then $R$ can be transformed into a
normalized proper scoring rule $\tilde{R}$ as follows.  As an
intermediate step, define $R'(y,x) = R(y,x) + \beta'(x)$ with
$\beta'(0) = -R(0,0)$ and $\beta'(1) = -R(1,1)$. Since $R$ is strictly
proper, so is $R'$, and both the maximum and the minimum must be taken
for $y \in \{0,1\}$. In particular, it must hold that both
$0=R'(0,0) > R'(1,0)$ and $0=R'(1,1) > R'(0,1)$. Let
$r_0:= R'(0,0) - R'(1,0)$ and $r_1:= R'(1,1) - R'(0,1)$ be the
intervals (``ranges'') of $R'$ for $X=0$ and $X=1$, respectively. Then
$\tilde{R}(y,x):= \frac{1}{\max(r_0,r_1)} R'(y,x) + 1$ is a normalized
scoring rule.

\section{Proper Scoring Rule Selection Violates Incentive Compatibility} \label{sec:mbr-fails}

  Let $R$ be any strictly proper scoring rule. Consider an instance with $m \geq 1$, and $n \geq 2$. Suppose that
  $\bm{p}_i=(0.5, \ldots, 0.5,0.8)$,\footnote{We instantiate a
    particular $\bm{p}_i$, but the example is not sensitive to this
    choice.} and consider joint distribution $D$ over $\bm{X}$ and
  $\bm{Y}_{-i}$ defined as follows.
	\begin{itemize}
	\item With probability 0.4, $\bm{X}=(0, \ldots, 0,1)$ and $\bm{Y}_j=(0.5, \ldots, 0.5, 0.8 + \frac{j}{10n})$ for all $j \neq i$.
	\item With probability 0.4, $\bm{X}=(1, \ldots, 1,1)$ and $\bm{Y}_j=(0.5, \ldots, 0.5, 0.8 + \frac{j}{10n})$ for all $j \neq i$.
	\item With probability 0.1, $\bm{X}=(0, \ldots, 0,0)$ and $\bm{Y}_j=(0.5, \ldots, 0.5, 0.8 + \frac{j}{10n})$ for all $j \neq i$.
	\item With probability 0.1, $\bm{X}=(1, \ldots, 1,0)$ and $\bm{Y}_j=(0.5, \ldots, 0.5, 0.8 + \frac{j}{10n})$ for all $j \neq i$.
	\end{itemize}
	Note in particular that $\mathbf{E}_{\bm{X} \sim D}\bigl[\bm{X}\bigr] = \bm{p}_i$, and that $0.8<Y_{j,m}\le 0.9$ with probability 1 for all $j \neq i$ 
	
	If forecaster $i$ reports $\bm{p}_i$, then all forecasters receive the same score on all events except event $m$. Forecaster $i$ receives the highest score, and is therefore selected by $\MPSR$, whenever $X_m=0$, which occurs with probability 0.2. That is, $\Pr\limits_{\bm{X}, \bm{Y}_{-i} \sim D}\bigl(\MPSR(\bm{Y}_1, \dots,
  \bm{p}_i, \dots,
  \bm{Y}_n,\bm{X}) = i\bigr)=0.2$.
 
 However, if forecaster i reports $\bm{y'}_i=(0.5,\ldots, 0.5, 1)$, then she is selected by $\MPSR$ whenever $X_m=1$, which occurs with probability 0.8. That is, $\Pr\limits_{\bm{X}, \bm{Y}_{-i} \sim D}\bigl(\MPSR(\bm{Y}_1, \dots,
  \bm{y'}_i, \dots,
  \bm{Y}_n,\bm{X}) = i\bigr)=0.8>0.2=\Pr\limits_{\bm{X}, \bm{Y}_{-i} \sim D}\bigl(\MPSR(\bm{Y}_1, \dots,
  \bm{p}_i, \dots,
  \bm{Y}_n,\bm{X}) = i\bigr)$, violating incentive compatibility.

\section{Proof of Theorem~\ref{thm:deterministic-not-truthful}}

Let $M$ be a deterministic and strictly incentive compatible
forecasting competition mechanism. Further, let $m \ge 1$, $n \ge 2$,
and observe that there are $|\cal{P}( [m])|= 2^m$ possible values of
the outcome vector $\bm{x}$. Consider forecaster $i$, and suppose that
every forecaster $j \neq i$ reports a probability $y_{j,k}=0.5$ for
every event $k$. We first use these fixed reports of agents $j \neq i$
to derive candidate misreports for agent $i$, and then again to define
an appropriate joint distribution $D$ that yields a violation of
strict incentive compatibility.

For any report $\bm{y}_i$, forecaster $i$ is selected as
the winner for some subset of possible event outcomes
$\cal{X} \subseteq \{ 0,1 \}^m$. Note that---since there are $2^m$
possible values of $\bm{x}$---there are
$|\cal{P}( \{ 0,1 \}^m)|=2^{2^m}$ possible subsets $\cal{X}$.

Consider then $2^{2^m}+1$ different possible reports of forecaster
$i$, denoted $\bm{y}_i^0, \bm{y}_i^1, \ldots, \bm{y}_i^{2^{2^m}}$, and
the corresponding subsets
$\cal{X}^0, \cal{X}^1, \ldots, \cal{X}^{2^{2^m}}$ of event outcomes
for which she is selected given these reports. By the pigeonhole
principle there must exist $r, s \in \{0, \dots, 2^{2^m}\}$ with
$r \neq s$ such that $\cal{X}^r=\cal{X}^s$. That is, forecaster $i$ is
selected for exactly the same set of possible event outcomes
regardless of whether she reports $\bm{y}_i^r$ or $\bm{y}_i^s$.

We use this fact to illustrate a violation of strict incentive
compatibility. Define $D$ as follows: each event $k$ occurs with
probability equal to $y^r_{i,k}$ independent of other events, and
every forecaster $j \neq i$ reports a probability of 0.5 for every
event. Note that $\bm{p}_i=\bm{y}_i^r$. Then we have that
$\Pr\limits_{\bm{X}, \bm{Y}_{-i} \sim D}\bigl(M(\bm{Y}_1, \dots,
\bm{p}_i, \dots, \bm{Y}_n,\bm{X}) = i\bigr) = \Pr\limits_{\bm{X} \sim
  D}\bigl( \bm{X} \in \cal{X}^r \bigr) = \Pr\limits_{\bm{X} \sim
  D}\bigl( \bm{X} \in \cal{X}^s \bigr) = \Pr\limits_{\bm{X},
  \bm{Y}_{-i} \sim D}\bigl(M(\bm{Y}_1, \dots, \bm{y}_i^s, \dots,
\bm{Y}_n,\bm{X}) = i\bigr)$, violating strict incentive
compatibility.\Halmos

\section{Multiplicatively Normalizing Scores From Proper Scoring Rules Violates Truthfulness}
\label{sec:strawman-fails}

  Let $n=2$, $m=1$, and suppose $p_1=0.5$. Let distribution $D$ over
  $X$ and $Y_2$ be defined as follows. With probability 0.5, $Y_2=1$
  and $X=0$, and with probability 0.5, $Y_2=1$ and $X=1$. Observe that
  $\mathbf{E}_{X \sim D}\bigl[X\bigr] = p_1$. If forecaster 1 reports
  $p_1$, then she is selected with probability
  $R_q(0.5,1)/\bigl(R_q(0.5,1)+R_q(1,1)\bigr) = 0.75/1.75 = 3/7$ when
  $X=1$, and $R_q(0.5,0)/\bigl(R_q(0.5,0)+R_q(1,0)\bigr) = 1$ when
  $X=0$.  That is,
  $\Pr\limits_{X, Y_2 \sim D}\bigl(M(p_1, Y_2,X) = 1\bigr)=5/7 \approx
  0.71$. If forecaster 1 instead reports $y'_1 = 0.8$, then she is
  selected with probability
  $R_q(0.8,1)/\bigl(R_q(0.8,1)+R_q(1,1)\bigr) = 0.96/1.96 = 24/49$
  when $X=1$, and $R_q(0.8,0)/\bigl(R_q(0.8,0)+R_q(1,0)\bigr) = 1$
  when $X=0$. Her probability of being selected has increased to
  $\Pr\limits_{X, Y_2 \sim D}\bigl(M(y'_1, Y_2,X) = 1\bigr)=73/98
  \approx 0.74$, violating truthfulness.

\section{Proof of Theorem~\ref{thm:single-event-truthfulness}}

To show strict truthfulness of $\MELFR$ for $m=1$, we show that
reporting $y_i = p_i$ maximizes forecaster $i$'s probability of being
selected for any joint distribution over outcomes $X$ and reports
$Y_{-i}$:
\begin{equation*}
  \begin{split}
    &\,\argmax_{y_i} \Pr\limits_{X, Y_{-i} \sim D}\bigl(\MELFR(Y_1,
    \dots, y_i, \dots,
    Y_n,X) = i\bigr)\\
    = &\,\argmax_{y_i} \E_{X,Y_{-i} \sim D} \left[f_i(Y_1, \dots, y_i, \dots, Y_n, X)\right]\\
    = &\,\argmax_{y_i} \E_{X,Y_{-i} \sim D} \left[\frac{1}{n} +
      \frac{1}{n} \biggl(R\bigl(y_i,X\bigr)-\frac{1}{n-1} \sum_{j
        \not= i} R\bigl(Y_j,X\bigr)\biggr)\right]\\
    = &\,\argmax_{y_i} \E_{X,Y_{-i} \sim D} \left[R\bigl(y_i,X\bigr)\right] = p_i
\end{split}
\end{equation*}
The last line follows from linearity of expectation and from $R$ being a strictly proper scoring
rule. \Halmos

\section{Proof of Theorem~\ref{thm:a-elf-truthfulness}}

\begin{equation*}
  \begin{split}
    &\,\argmax_{\bm{y}_i} \Pr\limits_{\bm{X}, \bm{Y}_{-i} \sim D}\bigl(\MTFSR(\bm{Y}_1,
    \dots, \bm{y}_i, \dots,
    \bm{Y}_n,\bm{X}) = i\bigr)\\
    = &\,\argmax_{\bm{y}_i} \E_{\bm{X}, \bm{Y}_{-i} \sim D} \left[g_i(\bm{Y}_1,
    \dots, \bm{y}_i, \dots,
    \bm{Y}_n,\bm{X})\right]\\
    = &\,\argmax_{\bm{y}_i} \E_{\bm{X}, \bm{Y}_{-i} \sim D} \left[\frac{1}{m}\sum_{k=1}^m \left( \frac{1}{n} + \frac{1}{n}
\biggl(R\bigl(\yik,X_k\bigr)-\frac{1}{n-1} \sum_{j \not= i}
R\bigl(Y_{j,k},X_k\bigr)\biggr)\right)\right]\\
    = &\,\argmax_{\bm{y}_i} \E_{\bm{X}, \bm{Y}_{-i} \sim D} \left[\sum_{k=1}^m \biggl(R\bigl(\yik,X_k\bigr)-\frac{1}{n-1} \sum_{j \not= i}
R\bigl(Y_{j,k},X_k\bigr)\biggr)\right]\\
    = &\,\argmax_{\bm{y}_i} \E_{\bm{X}, \bm{Y}_{-i} \sim D} \left[\sum_{k=1}^m R\bigl(y_{i,k},X_k\bigr)\right] = \bm{p}_i
\end{split}
\end{equation*}
\Halmos

\section{Proof of Proposition~\ref{prop:single-event-accuracy}}

The statement follows directly from the definition of $\MTFSR$.
\begin{equation*}
  \begin{split}
    \Pr\limits_{\bm{X} \sim \theta}\bigl(\MTFSR\bigl(\bm{y}_1, \dots,
  \bm{y}_n,\bm{X}\bigr) = i\bigr) &= \E_{\bm{X} \sim \theta} \left[ \frac{1}{m}\sum_{k=1}^m \Biggl( \frac{1}{n} + \frac{1}{n}
\biggl(R\bigl(\yik,X_k\bigr)-\frac{1}{n-1} \sum_{j \not= i}
R\bigl(y_{j,k},X_k\bigr)\biggr) \Biggr) \right]\\
&= \frac{1}{n}+\frac{1}{n}\biggl(\E_{\bm{X} \sim \theta}
\frac{1}{m}\sum_{k=1}^m \biggl[R\bigl(\yik,X_k\bigr)\biggr] - \E_{\bm{X} \sim \theta}
\frac{1}{m}\sum_{k=1}^m \biggl[\frac{1}{n-1} \sum_{j \not= i}
R\bigl(y_{j,k},X_k\bigr)\biggr]\biggr)\\
&= \frac{1}{n}+\frac{1}{n}\biggl(R(\bm{y}_i, \theta) - \frac{1}{n-1}
    \sum_{j \not= i} R(\bm{y}_j, \theta)\biggr)
\end{split}
\end{equation*}

\section{Proof of Theorem~\ref{thm:cant-do-better}}

Our proof of Theorem~\ref{thm:cant-do-better} proceeds in two parts. In the first part, we exploit the connection between wagering mechanisms and forecasting competition mechanisms to narrow down the particular form that any smooth, anonymous, strictly truthful forecasting competition mechanism must take. This form is parameterized by the choice of proper scoring rule $R$. In the second part of the proof, we show that using any normalized proper scoring rule different from the one used to define accuracy must violate rank accuracy.
Since we are considering only a single event $X$, for this proof we will slightly abuse notation and use $\theta$ to denote a single probability rather than a joint distribution.

{\bf Part 1.} We begin by formally introducing wagering mechanisms. A wagering mechanism $\Pi = (\Pi_i)_{i \in [n]}$ is a set of functions $\Pi_i$, each of which takes as input the forecasters' reports $\bm{y} = (y_1, \ldots, y_n) \in [0,1]^n$, a vector of wagers $\bm{\omega} = (\omega_1, \ldots, \omega_n) \in \mathbb{R}_{\ge 0}^n$, and the event outcome $x \in \{ 0, 1 \}$, and outputs a payment to forecaster $i$, $\Pi_i(\bm{y}, \bm{\omega}, x) \ge 0$. For our analysis, it will be sufficient to restrict ourselves to wagering mechanisms that only accept the vector of wagers $\bm{\omega}=(1/n, \ldots, 1/n)$. We refer to the resulting mechanisms as \emph{equal-wager wagering mechanisms},\footnote{We note that equal-wager wagering mechanisms can be equivalently expressed as \emph{Competitive Scoring Rules}~\citep{kilgour-gerchak:2004}.} and denote the payments $\Pi_i(y_1, \ldots, y_n,x)$, omitting the (non-)dependence on $\bm{\omega}$.

The following definitions are standard in the wagering mechanism literature.

\begin{definition}
An equal-wager wagering mechanism $\Pi$ is \emph{budget balanced} if, for all reports $y_1, \ldots, y_n \in [0,1]$ and outcomes $x \in \{ 0, 1 \}$, it holds that $\sum_{i=1}^n \Pi_i(y_1, \ldots, y_n,x)=1$. That is, the sum of payments from the mechanism equals the sum of agents' wagers. 
\end{definition} 

\begin{definition}
An equal-wager wagering mechanism $\Pi$ is \emph{strictly incentive compatible under immutable beliefs} if, for all $p_i$, all reports $y_i \neq p_i$, and all $y_j \in [0,1]$ for $j \neq i$, it holds that $\E_{X \sim p_i} \Pi_i(y_1, \ldots, y_i, \ldots, y_n,X) < \E_{X \sim p_i} \Pi_i(y_1, \ldots, p_i, \ldots, y_n, X)$. That is, truthfully reporting their subjective probability maximizes a forecaster's expected payment, given the reports of the other forecasters.
\end{definition}

\begin{definition}
An equal-wager wagering mechanism $\Pi$ is \emph{normal} if, for all probabilities $\theta \in [0,1]$, all reports $y_1, \ldots, y_n \in [0,1]$ and all $y'_i \in [0,1]$, if $\E_{X \sim \theta} \Pi_i(y_1, \ldots, y_i, \ldots, y_n,X) < \E_{X \sim \theta} \Pi_i(y_1, \ldots, y'_i, \ldots, y_n, X)$ then $\E_{X \sim \theta} \Pi_j(y_1, \ldots, y_i, \ldots, y_n,X) \ge \E_{X \sim \theta} \Pi_j(y_1, \ldots, y'_i, \ldots, y_n, X)$ for all $j \neq i$. That is, if a forecaster $i$ changes her report yielding a change $\epsilon_i$ in her expected payment, the change in expected payments of all other forecasters $\epsilon_j$ is null or has the opposite sign of $\epsilon_i$.
\end{definition}

\begin{definition}
An equal-wager wagering mechanism $\Pi$ is \emph{anonymous} if for any permutation $\sigma$ of $[n]$, any forecaster $i$, and any outcome $x$, it holds that $\Pi_i(y_1, \ldots, y_n,x) = \Pi_{\sigma(i)}(y_{\sigma^{-1}(1)}, \ldots, y_{\sigma^{-1}(n)},x)$. That is, the payouts do not depend on the identities of the agents.
\end{definition}

It will be useful to define smoothness for wagering mechanisms and proper scoring rules.

\begin{definition}
	An equal-wager wagering mechanism is \emph{smooth} if, for all $i \in [n]$, $\Pi_i$ is twice continuously differentiable with respect to each report $y_j$, $j \in [n]$. A proper scoring rule $R$ is smooth if it is twice continuously differentiable with respect to the report $y$.
\end{definition}

Our first lemma provides a formal relationship between budget-balanced equal-wager wagering mechanisms and forecasting competition mechanisms. 

\begin{definition}
	\label{def:corresponding}
	Given a forecasting competition mechanism $M$, define the \emph{corresponding} equal-wager wagering mechanism by $\Pi_i^M(y_1, \ldots, y_n ,x)= \text{Pr}(M(y_1, \ldots, y_n,x)=i) \ge 0$ for all $i \in [n]$.
\end{definition}

\begin{lemma}
	\label{lem:one-to-one}
	If a forecasting competition mechanism $M$ is strictly incentive compatible, anonymous, and smooth, then the corresponding equal-wager wagering mechanism $\Pi^M$ is budget-balanced, strictly incentive compatible for immutable beliefs, anonymous, and smooth.
\end{lemma}

\proof{Proof.}
	Consider a strictly incentive compatible and anonymous forecasting competition mechanism $M$ and the corresponding equal-wager wagering mechanism $\Pi^M$. 
	
	For budget balance, note that $\sum_{i=1}^n \Pi_i^M(y_1, \ldots, y_n,x)=\sum_{i=1}^n \text{Pr}(M(y_1, \ldots, y_n,x)=i)=1$, where the latter equality follows from the fact that $M$ outputs a probability distribution over forecasters.
	
	For anonymity, we have $\Pi_i^M(y_1, \ldots, y_n,x) = \text{Pr}(M(y_1, \ldots, y_n,x)=i) = \text{Pr}(M(y_{\sigma^{-1}(1)}, \dots, y_{\sigma^{-1}(n)},x)=\sigma(i)) = \Pi_{\sigma(i)}^M(y_{\sigma^{-1}(1)}, \ldots, y_{\sigma^{-1}(n)},x)$.
	
	For strict incentive compatibility under immutable beliefs, for any $p_i$, reports $y_i \neq p_i$, and any $y_j \in [0,1]$ for $j \neq i$, we have 
	\begin{align*}
		\E_{X \sim p_i} \Pi_i^M(y_1, \ldots, y_i, \ldots, y_n,X) 
		&=\text{Pr}_{X \sim p_i}(M(y_1, \ldots,y_i, \ldots, y_n,X)=i)\\
		&< \text{Pr}_{X \sim p_i}(M(y_1, \ldots,p_i, \ldots, y_n,X)=i)\\
		&=\E_{X \sim p_i} \Pi_i^M(y_1, \ldots, p_i, \ldots, y_n, X)
	\end{align*} 
	Where the inequality follows from strict incentive compatibility of $M$, taking joint distribution $D$ to be such that $Y_j=y_j$ with probability 1, and $\E_{X \sim D}[X]=p_i$.
	
	Finally, smoothness of $\Pi^M$ follows directly from smoothness of $M$ and Definition~\ref{def:corresponding}.
	\Halmos
\endproof

Lambert et al.~\shortcite{lambert-et-al:2008a} show that any smooth
equal-wager wagering mechanism that is budget balanced, strictly
incentive compatible for immutable beliefs, normal, and anonymous must
have a particular form. We note that the versions of normality and
incentive compatibility for immutable beliefs that Lambert et
al.~\shortcite{lambert-et-al:2008a} define are slightly weaker than
the ones we use. In particular, Lambert et al.\ do not require that
incentive compatibility holds for forecasters with belief $p_i=0$ or
$p_i=1$ and normality is required only to hold for $\theta \in
(0,1)$. The following statement still holds for our versions of these
properties since the mechanisms that satisfy our conditions are a
subset of the mechanisms that satisfy theirs.

\begin{lemma}[Lemma 4, Lambert et al.~\shortcite{lambert-et-al:2008a} (restated)]
	\label{thm:lambert}
  For any $n \ge 2$, if a smooth\footnote{Lambert et al. restrict attention to smooth wagering mechanisms, so this condition does not explicitly appear in their lemma statement.} equal-wager wagering mechanism $\Pi$ is budget balanced, strictly incentive compatible for immutable beliefs, anonymous and
  normal then there exists a smooth strictly proper scoring
  rule $R$ such that 
	\begin{equation}
		\label{eq:kg}
	 \Pi_i(y_1, \ldots, y_n,x) = \frac{1}{n} + R(y_i,x) - \frac{1}{n-1} \sum_{j \not=
        i} R(y_j,x).
		\end{equation}
\end{lemma}

The following lemma incorporates two observations about Lemma~\ref{thm:lambert}. First, $R$ must be bounded to guarantee non-negative payouts as required by the definition of a wagering mechanism. Second, when restricted to $n=2$, normality is implied by budget balance. 

\begin{lemma}
		\label{thm:lambert-n-2}
	  For $n=2$, if an equal-wager wagering mechanism is budget balanced, strictly incentive compatible for immutable beliefs, anonymous, and smooth, then there exists a smooth strictly proper scoring
	  rule $R \in [0,1]$ such that 
		\begin{equation} 
			\label{eq:kg-2-agents}
			\Pi_i(y_1,y_2,x) = \frac{1}{2} + \frac{1}{2} \left( R(y_i,x) - R(y_{3-i},x) \right).
		\end{equation}
	\end{lemma}
	
\proof{Proof.}
	When $n=2$, budget balance implies that $\Pi_1(y_1,y_2,x)=1-\Pi_2(y_1,y_2,x)$ for all $y_1,y_2 \in [0,1]$ and all $x \in \{ 0,1 \}$. Taking the expectation over possible outcomes yields $\E_{X \sim \theta} \Pi_1(y_1,y_2,X)=1-\E_{X \sim \theta}\Pi_2(y_1,y_2,X)$. In particular, any change in the expected payment to forecaster $i$ is exactly offset by the change in expected payment to forecaster $3-i$. Therefore, normality is implied by budget balance.
	
	Boundedness of $R$ follows from Lemma~\ref{thm:lambert} and the definition of a wagering mechanism. By the constraint that $0 \le \Pi_i(y_1, y_2, x) \le 1$, where $\Pi_i$ is defined as in Lemma~\ref{thm:lambert}, it must be the case that $|R(y_i,x) - R(y_{3-i},x)| \le 0.5$ for all $y_1,y_2,x$. We can therefore define $R'(y,x)$ by $R'(y,x)=2(R(y,x)+\beta(x))$, where $\beta(0)=-R(1,0)$ and $\beta(1)=-R(0,1)$. 
	$R'$ is derived from $R$ by a positive affine transformation, and therefore inherits strict properness from $R$.
	Note that the minimum value of $R'$ is $R'(0,1)=R'(1,0)=0$ and the maximum value is either $R'(0,0)= 2(R(0,0)-R(1,0)) \le 1$ or $R'(1,1) = 2(R(1,1)-R(0,1)) \le 1$, and so $R'$ is bounded in $[0,1]$. Further, plugging $R'$ into Equation~\ref{eq:kg-2-agents} yields exactly Equation~\ref{eq:kg}.
	\Halmos
\endproof

We can now characterize the form that any strictly incentive-compatible, anonymous, and smooth forecasting competition mechanism must have.

\begin{lemma}
		\label{lem:form-2-agents}
	  For $n=2$, if a forecasting competition mechanism $M$ is strictly incentive compatible, anonymous, and smooth, then there exists a smooth strictly proper scoring
	  rule $R(y,x) \in [0, 1]$ such that for all $i \in \{ 1,2 \}$
		\begin{equation*}
			\Pr (M(y_1,y_2,x)=i) = \frac{1}{2} + \frac{1}{2} \bigl( R(y_i,x) - R(y_{3-i},x) \bigr).
		\end{equation*}
	\end{lemma}
	
\proof{Proof.}
Let $M$ be a strictly incentive compatible, anonymous, and smooth forecasting competition mechanism. Then, by Lemma~\ref{lem:one-to-one}, the corresponding equal-wager wagering mechanism $\Pi^M$ is budget balanced, strictly incentive compatible for immutable beliefs, anonymous, and smooth. Therefore, by Lemma~\ref{thm:lambert-n-2}, there must exist a smooth strictly proper scoring rule $R \in [0,1]$ such that for all $i \in \{ 1,2 \}$
\[ \Pi_i^M(y_1,y_2,x) = \frac{1}{2} + \frac{1}{2} \left( R(y_i,x) - R(y_{3-i},x) \right). \]
By Definition~\ref{def:corresponding}, this implies that for all $i \in \{ 1,2 \}$
\[ \text{Pr}(M(y_1,y_2,x)=i) = \frac{1}{2} + \frac{1}{2} \bigl( R(y_i,x) - R(y_{3-i},x) \bigr), \]
which is the desired result.
\Halmos
\endproof
 
We have now established the form that any strictly incentive compatible, smooth, and anonymous forecasting competition mechanism $M$ must have for $n=2$. In particular, $M$ is equivalent to $\MELFR$ for some smooth, bounded proper scoring rule $R$. Further, we show that $R$ can always be represented by a differentiable convex function $G$.

\begin{lemma}
	\label{lem:savage}
	Let $R$ be a smooth strictly proper scoring rule. There exists a strictly convex, differentiable function $G: [0,1] \to \mathbb{R}$ with 
	\[ R(y,\theta)= G(y) + dG(y) \cdot (\theta-y), \]
where $\theta \in [0,1]$ and $dG(y)$ is the derivative of $G$ at $y$. Furthermore, $G(y)$ is the expected score for reporting $y=\theta$. Every $R$ defines a unique $G$ and every $G$ defines a unique $R$.
\end{lemma}

\proof{Proof.}
It is well known that every strictly proper scoring rule can be expressed as $R(y,\theta)= G(y) + dG(y) \cdot (\theta-y)$ for some strictly convex function $G$, where $dG(y)$ is a subgradient of $G$ at $y$~\citep{mccarthy1956measures,savage:1971,schervish1989general,gneiting-raftery:2007}. Observe that setting $y=\theta$ yields expected score $G(y)$, and it immediately follows that every $R$ defines a unique $G$.

Let $R$ be smooth (and, in particular, continuous). Suppose for the sake of contradiction that the convex function $G$ associated with $R$ is not differentiable at some $y' \in [0,1]$. 
That is, the left and right derivatives of $G$ at $y'$ ($d_-G(y')$ and $d_+G(y')$ respectively) are not equal. Note that convexity implies that $d_-G(y') \le d_+G(y')$, so the fact that the left and right derivatives are not equal yields $d_-G(y') < d_+G(y')$. We therefore have $\lim_{\epsilon \to 0^+} R(y'-\epsilon,1)=G(y') + d_-G(y') \cdot (1-y')<G(y') + d_+G(y') \cdot (1-y') = \lim_{\epsilon \to 0^+} R(y'+\epsilon,1)$, violating continuity of $R$ at $y'$ for $\theta=1$, a contradiction to smoothness of $R$.
Further, note that differentiability of $G$ implies a unique scoring rule $R$.
\Halmos
\endproof
\medskip

{\bf Part 2:} The remainder of the proof is devoted to comparing the behavior of $\MELFR$ for different choices of smooth proper scoring rule $R$. 
We will require the notion of equivalent scoring rules. A proper scoring rule $R$ is equivalent to another proper scoring rule $R'$ if $R$ can be obtained from $R'$ by a positive affine transformation.

\begin{definition}
  Proper scoring rules $R$ and $R'$ are \emph{equivalent} if
  and only if $R'(y,x) = \alpha R(y,x) + \beta(x)$ for some
  $\alpha > 0$ and $\beta(x) \in \mathbb{R}$ for $x \in \{ 0, 1 \}$.
\end{definition}

This definition partitions the space of proper scoring rules into
equivalence classes. It will be useful to define the \emph{canonical
  form} of a scoring rule $R$ as a convenient representative of each
class. In particular, the canonical form ensures that every perfect
forecast of a sure event obtains a score of 1 and that the minimum
expected score of a perfect forecast is 0.

\begin{definition}\label{def:canonical-form}
  Let $R$ and $R'$ be strictly proper scoring rules. We say that $R'$
  is the \emph{canonical form} of $R$ if $R'$ and $R$ are equivalent,
  and $R'(0,0) = R'(1,1) = 1$ as well as
  $\min_\theta R'(\theta,\theta) = 0$ for some $\theta \in (0,1)$.
\end{definition}

\begin{lemma} \label{lem:equivalent-statements} For any smooth proper scoring
  rule $R$, there exists a canonical form $R'$.
\end{lemma}

\proof{Proof.} It is sufficient to show that any strictly proper
scoring rule $R$ can be brought into canonical form through one
particular positive-affine transformation.

To transform any proper scoring rule $R$ into its canonical
form, we first define linear function $f(x)$ for $x \in \{0,1\}$ such
that, when added to $R(y,x)$, every perfect forecast of a sure event
obtains a score of 0. That is, $f(0) :=
- R(0,0)$ and $f(1) :=
- R(1,1)$.
In a second step, we are multiplying $R(y,x) + f(x)$ by
$\alpha := \frac{1}{-\min_{\theta} \E_{X \sim \theta}\left[R(\theta,X)
    + f(X)\right]}$ such that its minimum expected score of a perfect
forecast is -1. Note that $\alpha > 0$ since
$\min_{\theta} \E_{X \sim \theta}\left[R(\theta,X) + f(X)\right] < 0$
because $R(0,0) + f(0) = 0$ and $R(1,1) + f(1) = 0$ by design of
$f(x)$ and because of strict convexity of the expected score
function. Finally,
we add a constant $1$ to $R$, resulting in
$\alpha\bigl(R(y,x) + f(x)\bigr)+1$.  \Halmos \endproof

It immediately follows from Definition~\ref{def:canonical-form} and
Lemma~\ref{lem:equivalent-statements} that if two proper scoring rules
have the same canonical form, then they are equivalent. In order to prove our key result, we
require a technical lemma.

\begin{lemma}
	\label{lem:steeper}
	Let $f,g:[0,1] \to \mathbb{R}$ be differentiable, strictly convex functions. Additionally, suppose that $f$ is strictly decreasing, $f(0)=g(0)=1$ and that there exists a $\bar{t} \in (0,1]$ for which $f(\bar{t})<g(\bar{t})$. Then there must exist a $t' \in (0,\bar{t}]$ for which $f(t')<g(t')$ and $d(f(t'))<d(g(t'))$.
\end{lemma}

\proof{Proof.}
	Let $t^* = \text{sup} \{ t \in [0,\bar{t}] : f(t) \ge g(t) \}$. We are guaranteed that $t^*$ is well-defined because $f(0)=g(0)$ so we are taking a supremum over a non-empty set. Further, it is easy to see that $f(t^*) = g(t^*)$ and that $f(t)<g(t)$ for all $t \in (t^*, \bar{t}]$. Suppose for contradiction that $d(f(t)) \ge d(g(t))$ for all $t \in (t^*, \bar{t}]$. This would imply that $f(\bar{t}) \ge g(\bar{t})$, contradicting the assumption of the lemma. Therefore, there must exist a $t' \in (t^*, \bar{t}]$ with $d(f(t'))<d(g(t'))$.
	\Halmos
\endproof

Finally, we show  
that two smooth proper scoring rules $R$ and $R'$ are equivalent if and only if they always agree on the relative accuracy of forecasters.

\begin{lemma}
	\label{lem:different-rankings}
Smooth proper scoring rules $R$ and $R'$ are equivalent if and only if $R'(y_1,\theta) > R'(y_2,\theta) \Leftrightarrow R(y_1,\theta) > R(y_2,\theta)$ for all $y_1, y_2, \theta \in [0,1]$.
\end{lemma} 

\proof{Proof.}
We first prove the forward direction.
  Suppose that $R$ and $R'$ are equivalent, i.e.,
  $R'(y,x) = \alpha R(y,x) + \beta(x)$ for some $\alpha > 0$ and
  $\beta(x) \in \mathbb{R}$.
  Then,
  $R'(y_1,\theta) > R'(y_2,\theta) \Leftrightarrow \mathbf{E}_{X \sim \theta}\bigl[ \alpha R(y_1,X) + \beta(x) \bigr] > \mathbf{E}_{X \sim \theta}\bigl[ \alpha
  R(y_2,X) + \beta(X) \bigr] \Leftrightarrow \mathbf{E}_{X \sim \theta}\bigl[ \alpha R(y_1,X) \bigr]  + \mathbf{E}_{X \sim \theta} \bigl[\beta(X) \bigr] > \mathbf{E}_{X \sim \theta} \bigl[\alpha
  R(y_2,X)\bigr] + \mathbf{E}_{X \sim \theta} \bigl[\beta(X)\bigr] \Leftrightarrow
  \mathbf{E}_{X \sim \theta} \bigl[\alpha R(y_1,X)\bigr] > \mathbf{E}_{X \sim \theta} \bigl[\alpha
  R(y_2,X)\bigr] \Leftrightarrow  R(y_1,\theta) > R(y_2,\theta)$. 
  
  For the backward direction, suppose that $R$ and $R'$ are not
  equivalent. Assume that $R$ and $R'$ are in their respective
  canonical forms (if not, we can convert them to canonical form
  without changing the way they rank forecasters).
  Note that smoothness of $R$ and $R'$ implies the existence of associated differentiable convex functions $G$ and $G'$, as per Lemma~\ref{lem:savage}. Since $R$ and $R'$ are in canonical form, $\min_\theta G(\theta)=\min_\theta G'(\theta)=0$, and $G(0)=G(1)=G'(0)=G'(1)=1$. Further, since $R$ and $R'$ are not equivalent, we know that $G \neq G'$. We treat two cases.
  
Case 1: Suppose that $\text{arg} \min_\theta G(\theta) = \text{arg} \min_\theta G'(\theta)$. However, because $G \neq G'$, there must exist a $y$ at which $G(y) \neq G'(y)$. Without loss of generality, suppose $G(y)<G'(y)$. 
For mathematical convenience, suppose that $y < \text{arg} \min_\theta G(\theta)$; the case in which $y>\text{arg} \min_\theta G(\theta)$ follows similarly.

By Lemma~\ref{lem:steeper}, taking $f = G$ and $g = G'$, there must exist a point $y_1 <y$ for which $0<G(y_1)<G'(y_1)$ and $d(G(y_1)) < d(G'(y_1))<0$. 
Set $y_2=\text{arg} \min_\theta G(\theta)$ equal to the point at which $G(y_2)=G'(y_2)=0$. Since $G$ and $G'$ are both differentiable, $d(G(y_2))=d(G'(y_2))=0$.
Finally, set $\theta$ so that $R(y_1,\theta)=0$. That is,
\[ G(y_1) + d(G(y_1))(\theta-y_1)=0. \]
Note that, since $G(y_1)>0$ and $d(G(y_1))<0$, we have $\theta > y_1$.
Then,
\begin{align*}
	R(y_1,\theta) &= G(y_1)+d(G(y_1))\cdot (\theta-y_1)\\
	&= 0\\
	&= G(y_2)+d(G(y_2))\cdot (\theta-y_2)\\
	&= R(y_2,\theta)
\end{align*}
But,
\begin{align*}
	R'(y_1,\theta) &= G'(y_1)+d(G'(y_1))\cdot (\theta-y_1)\\
	&> G(y_1)+d(G(y_1))\cdot (\theta-y_1)\\
	&= 0\\
	&= G'(y_2)+d(G'(y_2))\cdot (\theta-y_2)\\
	&= R'(y_2,\theta),
\end{align*}
so that forecasters 1 and 2 obtain the same expected score according to $R$, but forecaster 1 obtains higher expected score according to $R'$. In particular, $R$ and $R'$ disagree on the relative accuracy.

Case 2: Suppose that, without loss of generality, $\theta_{min} := \text{arg} \min_\theta G(\theta) < \text{arg} \min_{\theta} G'({\theta}) := \theta'_{min}$. In particular, $G(\theta_{min})=0<G'(\theta_{min})$, and $G(\theta'_{min})>0=G'(\theta'_{min})$. By Lemma~\ref{lem:steeper}, there must exist a $y_1 < \theta_{min}$  for which $G(y_1) < G'(y_1)$ and $d(G(y_1))<d(G'(y_1))<0$. Similarly, there must exist a $y_2>\theta'_{min}$ for which $G(y_2)>G'(y_2)$ and $0 < d(G(y_2))<d(G'(y_2))$. Let $\theta$ be such that $R$ gives the same expected score to both reports. That is,
\[ R(y_1,\theta)=G(y_1)+d(G(y_1))\cdot (\theta - y_1) = G(y_2)+d(G(y_2)) \cdot (\theta - y_2)=R(y_2,\theta). \] 
Note that, by strict convexity of $G$, it needs to hold that $\theta \in (y_1, y_2)$. For $R'$ we have 
\begin{align*}
	R'(y_1,\theta) &= G'(y_1)+d(G'(y_1)) \cdot (\theta - y_1)\\
	&> G(y_1)+d(G(y_1)) \cdot (\theta - y_1)\\
	&= G(y_2)+d(G(y_2)) \cdot (\theta - y_2)\\
	&> G'(y_2)+d(G'(y_2)) \cdot (\theta - y_2)\\
	&= R'(y_2,\theta),
\end{align*}
where the first and last equalities follow from Lemma~\ref{lem:savage}, the inequalities hold because $\theta \in (y_1, y_2)$, and the second equality follows from the definition of $\theta$. Again, forecasters 1 and 2 obtain the same expected score according to $R$, but forecaster 1 obtains higher expected score according to $R'$. This completes the backward direction.
\Halmos
\endproof

We can now complete the proof of Theorem~\ref{thm:cant-do-better}.
	
\proof{Proof of Theorem~\ref{thm:cant-do-better}} By
Lemma~\ref{lem:form-2-agents} and Lemma~\ref{lem:savage}, when $n=2$, any smooth,
anonymous, strictly incentive-compatible forecasting competition mechanism $M$ must
take the form of $\MELF{R'}$ for some smooth, bounded proper scoring
rule $R' \in [0,1]$ with associated differentiable convex function
$G'$. We complete the proof by showing that every forecasting competition
mechanism of this form either fails to be rank accurate with respect
to $R$, or has
$\Pr\limits_{X \sim \theta}\bigl(M(y_1, y_2,X) = 1\bigr) \le
\Pr\limits_{X \sim \theta}\bigl(\MELFmax(y_1, y_2,X) = 1\bigr)$ for
every $y_1, y_2, \theta \in [0,1]$ for which
$R(y_1, \theta)>R(y_2,\theta)$.
  
If $R'$ is not equivalent to $R$, then $\MELF{R'}$ is not rank accurate with respect to $R$ by Corollary~\ref{cor:rank-accuracy} and Lemma~\ref{lem:different-rankings}
If $R'$ is equivalent to $R$ then we have that $R'(y,x)=\alpha R(y,x)+\beta (x)$. We also know that $\tilde{R}(y,x)=\tilde{\alpha} R(y,x) +\tilde{\beta}(x)$, where $\tilde{\alpha} \ge \alpha$ (if $\tilde{\alpha} < \alpha$ then $R'$ is not bounded in $[0,1]$). Let 
$y_1, y_2, \theta \in [0,1]$ such that $R(y_1, \theta)>R(y_2,\theta)$. Then
\begin{align*}
	\Pr\limits_{X \sim \theta}\bigl(\MELFmax(y_1, y_2,X) = 1\bigr) &= \frac{1}{2}+\frac{1}{2}(\tilde{R}(y_1,\theta)-\tilde{R}(y_2,\theta))\\
	&=\frac{1}{2}+\frac{1}{2}(\tilde{\alpha} R(y_1,\theta)+\E_{X \sim \theta} [\tilde{\beta} (X)]-\tilde{\alpha} R(y_2,\theta)-\E_{X \sim \theta} [\tilde{\beta} (X)])\\
	&=\frac{1}{2}+\frac{1}{2}(\tilde{\alpha} R(y_1,\theta)-\tilde{\alpha} R(y_2,\theta)\\
	&\ge\frac{1}{2}+\frac{1}{2}(\alpha R(y_1,\theta)-\alpha R(y_2,\theta)\\
	&=\frac{1}{2}+\frac{1}{2}(\alpha R(y_1,\theta)+\E_{X \sim \theta} [\beta (X)]-\alpha R(y_2,\theta)-\E_{X \sim \theta} [\beta (X)])\\
	&=\Pr\limits_{X \sim \theta}\bigl(\MELF{R'}(y_1, y_2,X) = 1\bigr),
	\end{align*}
where the inequality follows from $\tilde{\alpha} \ge \alpha$ and $R(y_1, \theta)>R(y_2,\theta)$.
\Halmos
  \endproof
  
\section{Proof of Theorem~\ref{thm:unbounded-impossibility}}

We first make a basic observation about unbounded proper scoring rules. The proof then proceeds by leveraging Lemma~\ref{lem:form-2-agents}, which characterizes the form that any strictly incentive-compatible, anonymous, and smooth forecasting competition mechanism must take. Finally, it shows that no mechanism of that form can be rank accurate with respect to an unbounded proper scoring rule.

Let $R$ be an unbounded strictly proper scoring rule. First note that since $R$ is strictly proper, it must be the case that 
$R(0,1)<R(y,1)$ for any $y>0$ and, analogously, $R(1,0)<R(y,0)$ for any $y<1$. Therefore, since $R$ is unbounded (i.e., $R(y,x)=-\infty$ for some $y \in [0,1]$ and $x \in \{ 0,1 \}$), it must be the case that $R(0,1)=-\infty$ and/or $R(1,0)=-\infty$, and $R(y,x) \in \mathbb{R}$ for all $y \in (0,1)$ and $x \in \{ 0,1 \}$.
Suppose now that $R(0,1)=-\infty$. (The case with $R(1,0)=-\infty$ can be proven identically.)

Let $M$ be a strictly incentive-compatible, anonymous, and smooth forecasting competition mechanism.
By Lemma~\ref{lem:form-2-agents}, we know the form that $M$ must take for $n=2$. In particular, there must exist a smooth, bounded strictly proper scoring rule $R' \in [0,1]$ such that $M=\MELF{R'}$.
We now show that $\MELF{R'}$ is not rank accurate with respect to $R$. Fix $y \in (0,1)$ and let $\theta$ be such that $R'$ gives the same expected score to reports $0$ and $y$. That is,
\[ R'(0,\theta)=G(0)+d(G(0)) \cdot \theta = G(y)+d(G(y)) \cdot (\theta-y)=R'(y,\theta), \]
where $G$ is the convex function associated with $R'$~\citep{savage:1971}. Note that $\theta \in (0,y)$ by strict convexity of $G$ and the fact that $d(G(0)) \in \mathbb{R}$ and $d(G(y)) \in \mathbb{R}$ (which is implied by boundedness of $R'$). Further, since $R'$ gives the same expected score to reports $0$ and $y$, if forecaster 1 reports $y_1=0$ and forecaster 2 reports $y_2=y$, we have 
\[ \Pr\limits_{X \sim \theta}\bigl(\MELF{R'}(y_1, y_2,X) = 1\bigr) = \Pr\limits_{X \sim \theta}\bigl(\MELF{R'}(y_1, y_2,X) = 2\bigr). \]

However, $R(y_1,\theta)=R(0,\theta)=\theta \cdot R(0,1)+(1-\theta) \cdot R(0,0)=\theta \cdot (-\infty)+(1-\theta) \cdot R(0,0)=-\infty$ and $R(y_2,\theta)=R(y,\theta) \in \mathbb{R}$. 
Therefore, $\MELF{R'}$ is not rank accurate with respect to $R$. 
  
\section{Proof of Proposition~\ref{thm:no-truthful-and-limit-accurate-mechanisms}}
\label{sec:proof-no-truthful-and-limit-accurate-mechanisms}
  
  \proof{Proof.} Let $n=2$ with $\bm{y}_1 = (0.4, \dots, 0.4)$ and
  $\bm{y}_2 = (0.6, \dots, 0.6)$. Let $R$ be the strictly proper scoring
  rule that defines accuracy. Now suppose $M$ is a limit accurate
  forecasting competition mechanism and consider the following two cases
  with two different ``perfectly correlated'' joint distributions
  $\theta$ for which all $m$ outcomes are the same, i.e., either
  $X_k = 0$ for all $k$ or $X_k = 1$ for all $k$:

  \begin{enumerate}

  \item $\theta_k = 0.4$ for all $k$. Since $y_{1,k} = \theta_k$ and
    $y_{2,k} \neq \theta_k$ for all $k$, strict properness of $R$
    implies that forecaster $1$ is strictly more accurate. Hence, limit
    accuracy implies that there exists an $\underline{m_1}$ such
    that for all $m \ge m_1$, $M$ selects forecaster $1$ with probability at least
    $\pi = 0.7$.

  \item $\theta_k = 0.6$ for all $k$. Since $y_{2,k} = \theta_k$
    and $y_{1,k} \neq \theta_k$ for all $k$, strict properness of $R$
    implies that forecaster $2$ is strictly more accurate. 
    Hence, limit accuracy implies that there exists an $\underline{m_2}$ such
      that for all $m \ge m_2$, $M$ selects forecaster $2$ with probability at least
      $\pi = 0.7$.
  \end{enumerate}

  Now let $m = \max(\underline{m_1},\underline{m_2})$ be the number of
  events. Since both $\theta$ are ``perfectly correlated,'' the outcome
  vector is either $\bm{x} = (0, \dots, 0)$ or $\bm{x} = (1, \dots, 1)$,
  and so it is sufficient to consider whom $M$ selects given each of
  these. Let $q_{1|\bm{0}}$ and $q_{1|\bm{1}}$ be the probabilities that
  $M$ selects forecaster $1$ given $\bm{x} = (0, \dots, 0)$ and
  $\bm{x} = (1, \dots, 1)$, respectively. From Case 1, it needs to hold
  that $0.4 \cdot q_{1|\bm{1}} + 0.6 \cdot q_{1|\bm{0}} > 0.7$ and from
  Case 2, it needs to hold that
  $0.6 \cdot (1-q_{1|\bm{1}}) + 0.4 \cdot (1-q_{1|\bm{0}}) > 0.7$. But
  this is impossible because the former implies that
  $q_{1|\bm{1}} > \frac{7}{4} - \frac{3}{2} q_{1|\bm{0}}$ and the latter
  implies that $q_{1|\bm{1}} < \frac{1}{2} - \frac{2}{3} q_{1|\bm{0}}$,
  with no $q_{1|\bm{0}},q_{1|\bm{1}} \in [0,1]$ satisfying both; and a
  contradiction that $M$ is limit accurate. \Halmos \endproof

\section{Proof of Theorem \ref{thm:multi-event-truthful}}

\proof{Proof.}
Without loss of generality, take the perspective of any forecaster $i \in [n]$ seeking to
maximize the probability of being selected. In reasoning about forecaster $i$'s probability of winning, she needs
to reason about the joint probability of the event winners vector
$(w_1, \dots, w_m)$, which is given by the vector of probability
distributions $(\bm{f}_1, \dots, \bm{f}_m)$, where each $\bm{f}_k$ is
the distribution over forecasters for event $k$. From forecaster $i$'s
perspective, each $\bm{f}_k$ is an instantiation of a random variable
$\bm{F}_k$, depending on her belief about $\bm{Y}_{-i}$ and
$\bm{X}$. Without any restrictions on $\bm{Y}_{-i}$ and
$\bm{X}$, these $\bm{F}_k$ can be
dependent even if---given instantiated
$(\bm{f}_1, \dots, \bm{f}_m)$---the draws of the event winners
themselves are independent by definition of the mechanism.
For belief independent joint distributions $D$ over outcomes $\bm{X}$
and reports $\bm{Y}_{-i}$, however, all random vectors
$(Y_{1,k}, \dots, Y_{i-1,k}, Y_{i+1,k}, \dots, Y_{n,k}, X_k)$ indexed
by $k$ are independent, so that all $\bm{F}_k$ are independent as
well. Consider now event $k$ and let $K' \in \mathcal{P}([m])$ be any
subset of event indices with $k \not\in K'$. By independence of
$\bm{F}_k$ for all $k$, changing forecaster $i$'s report on event $k$
does not affect the (joint) distribution of $\bm{F}_{K'}$. 

It is easy to see that increasing forecaster $i$'s expected
(subjective) winning probability for event $k$, $\mathbf{E}[F_{i,k}]$,
simultaneously decreases the expected winning probability
$\mathbf{E}[F_{j,k}]$ of every $j \neq i$. To see this, first observe
that, if $\mathbf{E}[F_{i,k}]$ increases, the sum of all other
forecasters' event winning probabilities needs to decrease by the same
amount since
$\mathbf{E}[F_{i,k}] + \sum_{j \neq i} \mathbf{E}[F_{j,k}] = 1$ for
all $k$. Second, by definition of $f_{i,k}$, any increase of
$\epsilon > 0$ in $\mathbf{E}[F_{i,k}]$ leads to a uniform decrease of
$\frac{\epsilon}{n-1}$ in each $\mathbf{E}[F_{j,k}]$ with
$j \neq i$. This
means that, since the $\bm{F}_k$ are independent, increasing
$\mathbf{E}[F_{i,k}]$ on event $k$ cannot decrease your probability of
winning overall.

It remains to be shown that increasing $\mathbf{E}[F_{i,k}]$ strictly
increases forecaster $i$'s probability of winning overall. To show
this, we need to show that there are situations, where event $k$ is
pivotal for winning overall and that these situations occur with
positive probability.
First, there exist event win outcomes
$w_1, \ldots, w_{k-1}, w_{k+1}, \ldots, w_m$ on the other $m-1$ events
such that $k$ is pivotal, i.e., winning or losing event $k$ changes
the probability of winning the prize. This is the case if and only if,
without event $k$, some forecaster $j \neq i$ won most events with
forecaster $i$ winning one fewer; or forecaster $i$ won most events
with at least one other forecaster $j \neq i$ having won exactly the
same number, or one event less than forecaster $i$. For example, with
$m$ odd, $m-1$ is even and forecasters $i$ and $j \neq i$ can each win
half of those events. Similarly, with $m$ even, $m-1$ is odd, and it
can be the case that forecaster $i$ wins
$\lfloor \frac{m-1}{2} \rfloor$ and $j$ wins
$\lceil \frac{m-1}{2} \rceil$. Second, these cases occur with positive
probability because we know that every $\mathbf{E}[F_{j,k}]$ for all
$j$ and all $k$ is strictly in between 0 and 1 by definition of
$f_{i,k}$ and $R \in [0,1)$. Hence, event $k$ is pivotal for
forecaster $i$ with positive probability, and reporting truthfully on
event $k$ strictly increases the probability of winning the prize.
\Halmos \endproof

\section{Proof of Theorem \ref{thm:multi-event-accuracy}}

The proof uses the one-sided version of Hoeffding's inequality
\citep{hoeffding:1963}, which we state here for convenience.

{\TheoremHeaderFont Theorem} {\bf (Hoeffding's inequality)}
\emph{Let
  $X_1, \ldots, X_m$ be independent random variables bounded by the
  interval $[0,1]$. Define $S_m = X_1 + \ldots +X_m$. Then
  \begin{equation*}
    \Pr\Bigl(S_m-\E \bigl[S_m\bigr] \ge t\Bigr) \le e^{-\frac{2t^2}{m}}.
  \end{equation*}
  and
  \begin{equation*}
    \Pr\Bigl(\E \bigl[S_m\bigr] - S_m \ge t\Bigr) \le e^{-\frac{2t^2}{m}}.
  \end{equation*}}

\proof{Proof.} Let
$w_{i,k} := \mathbbm{1}(w_k = i)$ indicate whether forecaster $i$ is
the event winner for event $k$, and let $W_{i,k}$ be the corresponding
random variable. Note that the reports $\bm{y}_1, \dots, \bm{y}_n$ are
fixed, so that the uncertainty is only about the event outcomes
$\bm{X}$. In particular, with $X_1, \dots, X_m$ independent,
$W_{i,1}, \dots, W_{i,m}$ are independent conditional on
$\bm{y}_1, \dots, \bm{y}_n$.

Let $z_i = \sum_{k = 1}^m w_{i,k}$ be the number of events won by
forecaster $i$. Furthermore, let $Z_i$ be the corresponding random
variable, so that
\begin{equation*}
\E_{\bm{X} \sim \theta} [Z_i] = \E_{\bm{X} \sim \theta} \left[\sum_{k=1}^m f_{i,k}\right],
\end{equation*}
where the latter expectation is taken over the outcomes, and the
former is taken over the outcomes and the randomness of the
lotteries.

To show limit accuracy, let $i$ be the most accurate forecaster with
$\Delta := \min_{j \neq i} \bigl(R(\bm{y}_i,\theta) -
R(\bm{y}_j,\theta) \bigr) > 0$ denoting the difference between the
expected scores of $i$ and the second-most accurate forecaster.
We first bound the difference between the
expected number of events won by $i$ and the expected number of events
won by some other forecaster $j \not= i$:
\begin{equation}\label{eq:exp-difference}
  \begin{split}
    &\E_{\bm{X} \sim \theta} \left[ Z_i \right] -
    \E_{\bm{X} \sim \theta} \left[ Z_j \right]
    = \E_{\bm{X} \sim \theta} \left[ \sum_{k=1}^m \bigl(f_{i,k}\! -\! f_{j,k}\bigr) \right]\\[0.2cm]
    = &\frac{\E\limits_{\bm{X} \sim \theta} \Bigl[ \sum_{k=1}^m
      \bigl(R(\yik, X_k)-R(y_{j,k}, X_k)\bigr) \Bigr]}{n-1}\\
    = &\frac{m \bigl(R(\bm{y}_i, \theta)-R(\bm{y}_j,
      \theta)\bigr)}{n-1} \ge
    \frac{m\Delta}{n-1}.\\[0.1cm]
  \end{split}
\end{equation}
The second equality follows from substituting the definition of
$f_{i,k}$ and simplifying, the third equality follows from rewriting
in terms of expected average score, and the inequality follows from
the definition of $\Delta$.

We now upper bound the probability that forecaster $j$ wins more
events than forecaster $i$. From Equation~\ref{eq:exp-difference}, if
$z_j \ge z_i$, then it holds that $\E [Z_i] - z_i \ge
\frac{m\Delta}{2(n-1)}$ or $z_j - \E [Z_j] \ge
\frac{m\Delta}{2(n-1)}$ (both may apply simultaneously). By Hoeffding's inequality,
  \begin{equation*}
    \Pr\biggl(\E \bigl[Z_i\bigr] - z_i \ge \frac{m\Delta}{2(n-1)} \biggr) \le e^{-\frac{m\Delta^2}{2(n-1)^2}},
  \end{equation*}
  and
  \begin{equation*}
    \Pr\biggl(z_j - \E \bigl[Z_j\bigr]\ge \frac{m\Delta}{2(n-1)} \biggr) \le e^{-\frac{m\Delta^2}{2(n-1)^2}},
  \end{equation*}
  Putting these together, we have
  \begin{equation*}
    \begin{split}
      \Pr(z_j \ge z_i) \le &\Pr\Biggl(\biggl(\E \bigl[Z_i\bigr] - z_i \ge
          \frac{m\Delta}{2(n-1)} \biggr) \cup \left( z_j - \E
          \bigl[Z_j\bigr] \ge \frac{m\Delta}{2(n-1)} \right)\Biggr)   \\[0.1cm]
      \le &\Pr\left(\E \bigl[Z_i\bigr] - z_i \ge
        \frac{m\Delta}{2(n-1)}\right) + \Pr\left(z_j - \E \bigl[Z_j\bigr] \ge \frac{m\Delta}{2(n-1)}\right)\\[0.1cm]
      \le &\,2e^{-\frac{m\Delta^2}{2(n-1)^2}}.
    \end{split}
  \end{equation*}
  
  Finally, we lower bound the probability that \elf selects forecaster
  $i$.
  \begin{equation*}
    \begin{split}
      \Pr_{\bm{X} \sim \theta}\Bigl(\MELFR\bigl(\bm{y}_1, \dots,
      \bm{y}_n,\bm{X}\bigr) = i\Bigr)
      = &\,1 - \sum_{j \not= i} \Pr_{\bm{X} \sim
        \theta}\Bigl(\MELFR\bigl(\bm{y}_1, \dots,
      \bm{y}_n,\bm{X}\bigr) = j\Bigr)\\
      \ge &\,1 - \sum_{j \not= i} \Pr_{\bm{X} \sim \theta}\Bigl( z_j \ge z_i \Bigr)\\
      \ge &\,1 - 2(n-1)e^{-\frac{m\Delta^2}{2(n-1)^2}},
    \end{split}
  \end{equation*}
  where the first transition holds because exactly one forecaster is
  selected and the second because $z_j \ge z_i$ is a necessary
  condition for forecaster $j$ to be selected by \elf. The final
  transition holds by plugging in the earlier inequality.
  In particular, for fixed $n$ and `accuracy gap' $\Delta$, for any
  $\pi \in [0,1)$, \ielf selects the best forecaster with probability at
  least $\pi$ if
  \begin{equation*}
    m \ge \frac{2(n-1)^2}{\Delta^2}\ln\left( \frac{2(n-1)}{1-\pi} \right),
  \end{equation*}
  which yields limit accuracy.
  \Halmos \endproof

\end{APPENDIX}

\bibliographystyle{informs2014} 
\bibliography{jens} 

\begin{thebibliography}{33}
\providecommand{\natexlab}[1]{#1}
\providecommand{\url}[1]{\texttt{#1}}
\providecommand{\urlprefix}{URL }

\bibitem[{Atanasov et~al.(2017)Atanasov, Rescober, Stone, Servan-Schreiber,
  Tetlock, Ungar, \protect\BIBand{} Mellers}]{atanasov-et-al:2017}
Atanasov P, Rescober P, Stone E, Servan-Schreiber E, Tetlock PE, Ungar L,
  Mellers B (2017) {Distilling the Wisdom of Crowds: Prediction Markets versus
  Prediction Polls}. \emph{Management Science} 63(3):691--706.

\bibitem[{Brier(1950)}]{brier:1950}
Brier GW (1950) {Verification of Forecasts Expressed in Terms of Probability}.
  \emph{{Monthly Weather Review}} 78(1):1--3.

\bibitem[{Chakraborty(2016)}]{chakraborty:2016}
Chakraborty A (2016) {How Companies Are Using Kaggle To Find The Best Machine
  Learning Talent}.
  \url{https://blog.udacity.com/2016/07/companies-kaggle-machine-learning-talent.html},
  [Online; accessed 24-December-2020].

\bibitem[{Chakraborty et~al.(2013)Chakraborty, Das, Lavoie, Magdon-Ismail,
  \protect\BIBand{} Naamad}]{chakraborty:2013}
Chakraborty M, Das S, Lavoie A, Magdon-Ismail M, Naamad Y (2013) Instructor
  rating markets. \emph{Twenty-Seventh AAAI Conference on Artificial
  Intelligence}.

\bibitem[{Cowgill \protect\BIBand{} Zitzewitz(2015)}]{cowgill-zitzewitz:2015}
Cowgill B, Zitzewitz E (2015) {Corporate Prediction Markets: Evidence from
  Google, Ford, and Firm X}. \emph{The Review of Economic Studies}
  82(4):1309--1341.

\bibitem[{Gneiting \protect\BIBand{} Raftery(2007)}]{gneiting-raftery:2007}
Gneiting T, Raftery AE (2007) {Strictly Proper Scoring Rules, Prediction, and
  Estimation}. \emph{Journal of the American Statistical Association}
  102:359--378.

\bibitem[{Good(1952)}]{good:1952}
Good IJ (1952) {Rational Decisions}. \emph{{Journal of the Royal Statistical
  Society. Series B}} 14(1):107--114.

\bibitem[{Grushka-Cockayne et~al.(2017)Grushka-Cockayne, Lichtendahl, Jose,
  \protect\BIBand{} Winkler}]{grushka-cockayne-et-al:2017}
Grushka-Cockayne Y, Lichtendahl KC, Jose VR, Winkler RL (2017) {Quantile
  Evaluation, Sensitivity to Bracketing, and Sharing Business Payoffs}.
  \emph{Operations Research} 65(3):557--836.

\bibitem[{Harris(2013)}]{harris:2013}
Harris D (2013) {Facebook is hiring a data scientist, but you'll have to fight
  for the job}.
  \url{https://gigaom.com/2013/08/30/facebook-is-hiring-a-data-scientist-but-youll-have-to-fight-for-the-job/},
  [Online; accessed 24-December-2020].

\bibitem[{Hoeffding(1963)}]{hoeffding:1963}
Hoeffding W (1963) {Probability Inequalities for Sums of Bounded Random
  Variables}. \emph{Journal of the American Statistical Association}
  58(301):13--30.

\bibitem[{Hossain \protect\BIBand{} Okui(2013)}]{hossain-okui:2013}
Hossain T, Okui R (2013) {The Binarized Scoring Rule}. \emph{The Review of
  Economic Studies} 80(3):984--1001.

\bibitem[{Jia et~al.(2017)Jia, Liu, Yu, \protect\BIBand{}
  Voida}]{jia-et-al:2017}
Jia Y, Liu Y, Yu X, Voida S (2017) Designing leaderboards for gamification:
  Perceived differences based on user ranking, application domain, and
  personality traits. \emph{Proceedings of the 2017 CHI conference on human
  factors in computing systems}, 1949--1960.

\bibitem[{Jose(2009)}]{jose:2009}
Jose VR (2009) {A Characterization for the Spherical Scoring Rule}.
  \emph{Theory and Decision} 66(3):263--281.

\bibitem[{Jose(2017)}]{jose:2017}
Jose VR (2017) {Percentage and Relative Error Measures in Forecast Evaluation}.
  \emph{Operations Research} 65(1):200--211.

\bibitem[{Kaggle(2017)}]{kaggle:2017}
Kaggle (2017) {March Machine Learning Mania, 1st Place Winner’s Interview:
  Andrew Landgraf}.
  \url{https://medium.com/kaggle-blog/march-machine-learning-mania-1st-place-winners-interview-}
  \url{andrew-landgraf-f18214efc659}, [Online; accessed 24-December-2020].

\bibitem[{Karni(2009)}]{karni:2009}
Karni E (2009) {A Mechanism for Eliciting Probabilities}. \emph{Econometrica}
  77(2):603--606.

\bibitem[{Kearns \protect\BIBand{} Vazirani(1994)}]{kearns:1994}
Kearns MJ, Vazirani UV (1994) \emph{{An Introduction to Computational Learning
  Theory}} (MIT press).

\bibitem[{Kilgour \protect\BIBand{} Gerchak(2004)}]{kilgour-gerchak:2004}
Kilgour DM, Gerchak Y (2004) {Elicitation of Probabilities Using Competitive
  Scoring Rules}. \emph{Decision Analysis} 1(2):108--113.

\bibitem[{Konrad(2009)}]{konrad:2009}
Konrad KA (2009) \emph{Strategy and Dynamics in Contests} (Oxford University
  Press).

\bibitem[{Lambert et~al.(2008)Lambert, Langford, Wortman, Chen, Reeves, Shoham,
  \protect\BIBand{} Pennock}]{lambert-et-al:2008a}
Lambert N, Langford J, Wortman J, Chen Y, Reeves D, Shoham Y, Pennock DM (2008)
  {Self-Financed Wagering Mechanisms for Forecasting}. \emph{Proceedings of the
  9th ACM Conference on Electronic Commerce (EC'08)}, 170--179 (ACM).

\bibitem[{Lambert(2011)}]{lambert:2011}
Lambert NS (2011) {Probability Elicitation for Agents with Arbitrary Risk
  Preferences}, {Working Paper}.

\bibitem[{Lichtendahl et~al.(2013)Lichtendahl, Grushka-Cockayne,
  \protect\BIBand{} Pfeifer}]{lichtendahl-et-al:2013}
Lichtendahl KC, Grushka-Cockayne Y, Pfeifer PE (2013) {The Wisdom of
  Competitive Crowds}. \emph{Operations Research} 61(6):1383--1398.

\bibitem[{Lichtendahl \protect\BIBand{}
  Winkler(2007)}]{lichtendahl-winkler:2007}
Lichtendahl KCJ, Winkler RL (2007) {Probability Elicitation, Scoring Rules, and
  Competition Among Forecasters}. \emph{Management Science} 53(11):1745--1755.

\bibitem[{Machina \protect\BIBand{} Schmeidler(1992)}]{machina-schmeidler:1992}
Machina MJ, Schmeidler D (1992) A more robust definition of subjective
  probability. \emph{Econometrica: Journal of the Econometric Society}
  745--780.

\bibitem[{McCarthy(1956)}]{mccarthy1956measures}
McCarthy J (1956) Measures of the value of information. \emph{Proceedings of
  the National Academy of Sciences} 42(9):654--655.

\bibitem[{Mellers et~al.(2014)Mellers, Ungar, Baron, Ramos, Gurcay, Fincher,
  Scott, Moore, Atanasov, Swift, Murray, Stone, \protect\BIBand{}
  Tetlock}]{mellers-et-al:2014}
Mellers B, Ungar L, Baron J, Ramos J, Gurcay B, Fincher K, Scott SE, Moore D,
  Atanasov P, Swift SA, Murray T, Stone E, Tetlock PE (2014) {Psychological
  Strategies for Winning a Geopolitical Forecasting Tournament}.
  \emph{Psychological Science} 25(5):1106--1115.

\bibitem[{Palley \protect\BIBand{} Soll(2019)}]{palley-soll:2019}
Palley AB, Soll JB (2019) {Extracting the Wisdom of Crowds When Information Is
  Shared}. \emph{Management Science} 65(5):1949--2443.

\bibitem[{Satop\"a\"a et~al.(2014)Satop\"a\"a, Baron, Foster, Mellers, Tetlock,
  \protect\BIBand{} Ungar}]{satopaa-et-al:2014}
Satop\"a\"a VA, Baron J, Foster DP, Mellers BA, Tetlock PE, Ungar LH (2014)
  {Combining multiple probability predictions using a simple logit model}.
  \emph{International Journal of Forecasting} 30(2):344--356.

\bibitem[{Savage(1971)}]{savage:1971}
Savage LJ (1971) {Elicitation of Personal Probabilities and Expectations}.
  \emph{Journal of the American Statistical Association} 66:783--801.

\bibitem[{Schervish et~al.(1989)}]{schervish1989general}
Schervish MJ, et~al. (1989) A general method for comparing probability
  assessors. \emph{The annals of statistics} 17(4):1856--1879.

\bibitem[{Servan-Schreiber et~al.(2004)Servan-Schreiber, Wolfers, Pennock,
  \protect\BIBand{} Galebach}]{servan-et-al:2004}
Servan-Schreiber E, Wolfers J, Pennock DM, Galebach B (2004) {Prediction
  Markets: Does Money Matter?} \emph{{Electronic Markets}} 14(3):243--251.

\bibitem[{Tetlock \protect\BIBand{} Gardner(2015)}]{tetlock-gardner:2015}
Tetlock PE, Gardner D (2015) \emph{{Superforecasting: The Art and Science of
  Prediction}} (New York, NY, USA: Crown Publishing Group).

\bibitem[{Witkowski et~al.(2018)Witkowski, Freeman, Wortman~Vaughan, Pennock,
  \protect\BIBand{} Krause}]{witkowski-et-al:2018}
Witkowski J, Freeman R, Wortman~Vaughan J, Pennock DM, Krause A (2018)
  {Incentive-Compatible Forecasting Competitions}. \emph{Proceedings of the
  32nd AAAI Conference on Artificial Intelligence (AAAI'18)}.

\end{thebibliography}

\end{document}